\pdfoutput=1
\documentclass{aa}  
\usepackage{graphicx}
\usepackage{pdflscape}
\usepackage{natbib}
\usepackage{txfonts}
\usepackage{float}
\usepackage{placeins}
\usepackage{dblfloatfix}
\usepackage{multirow,array} 
\usepackage{arydshln}
\usepackage{subcaption}
\usepackage[outdir=./]{epstopdf}
\usepackage[colorlinks=true,
            linkcolor=blue,
            urlcolor=blue,
            citecolor=blue]{hyperref}
\usepackage[nameinlink]{cleveref}
\Crefname{figure}{Fig.}{Figs.}

\begin{document} 

   \title{Optical spectroscopic and polarization properties of 2011 outburst of the recurrent nova T Pyxidis}


   \author{M. Pavana
          \inst{1,2}\thanks{e-mail: pavana@iiap.res.in}
          \and
          Ramya M Anche
          \inst{1}
          \and
          G. C. Anupama\inst{1}
          \and
          A. N. Ramaprakash\inst{3} \and  G. Selvakumar\inst{1}
          }

   \institute{Indian Institute of Astrophysics, 560 034 Bangalore, India\\
         \and
             Pondicherry University, R.V. Nagar, Kalapet, 605014, Puducherry, India\\
         \and
             Inter-University Center for Astronomy and Astrophysics, Post Bag 4, Ganeshkhind, 411 007 Pune, India\\
             }

   \date{Received; accepted}

  \abstract
   {}
   {To study the spectroscopic and ionized structural evolution of T Pyx during its 2011 outburst, and also study the variation in degree of polarization during its early phase.}
   {Optical spectroscopic data of this system obtained from day 1.28--2415.62 since discovery, and optical, broadband imaging polarimetric observations obtained from day 1.36--29.33 during the early phases of the outburst are used in the study. The physical conditions and the geometry of the ionized structure of the nova ejecta has been modelled for a few epochs using the photo-ionization code, CLOUDY in 1D and pyCloudy in 3D.}
   {The spectral evolution of the nova ejecta during its 2011 outburst is similar to that of the previous outbursts. The variation in the line profiles is seen very clearly in the early stages due to good coverage during this period. The line profiles vary from P-Cygni (narrower, deeper and sharper) to emission profiles which are broader and structured, which later become narrower and sharper in the late post-outburst phase. The average ejected mass is estimated to be $7.03\, \times\, 10^{-6}\, M_{\odot}$. The ionized structure of the ejecta is found to be a bipolar conical structure with equatorial rings, with a low inclination angle of $14.75^\circ\, \pm\, 0.65^\circ$.}
   {}


  \maketitle
%
\section{Introduction}
Recurrent Novae (RNe) are nova systems that have more than one recorded outburst. They play an important role in the understanding of nova outbursts because of their multiple eruptions. Observations during the several eruptions of these systems have led to a better understanding of nova explosions. So far, 10 Galactic RNe are known. These are sub-classified based on various observational features, out of which orbital period is one such feature \citep{gca08}. They can be divided into long and short period systems. The long period systems consist of the systems with red giants, RS Oph, T CrB, V3890 Sgr and V745 Sco. The short period systems are further divided into U Sco and T Pyx groups based on the outburst and quiescent properties. U Sco group consists of U Sco, V394 CrA, and V2487 Oph, and the T Pyx group consists of T Pyx, CI Aql and IM Nor. 

T Pyx is a well-known RN, whose eruptions were observed in 1890, 1902, 1920, 1945, 1966, and the most recent one in 2011. It was first discovered as a nova by H. Leavitt on the plates of Harvard Map from the 1902 outburst \citep{Due87}. The quiescence spectrum of T Pyx is dominated by the accretion disc spectrum \citep{gca08, Sel08}. The 2011 outburst was discovered by M. Linnolt at 13.0 V magnitude \citep{Sch13}, on April 14.29 UT (JD 2455665.79), and was well studied with multi-wavelength observations.

The spectroscopic and photometric properties of the 2011 eruption have been reported by \citet{Che11,Sho13,Sok13,deg14,Sur14,Jos14}. Using the light echoes method, \citet{Sok13} estimated the distance to T Pyx as 4.8 kpc while \citet{Sch18} has estimated the distance to be $\sim$ 3.1 kpc using the parallax data from Gaia. The optical and IR spectra indicate T Pyx morphed from a He/N to Fe II type within the first few days, on its rise to the maximum, and transitioned to He/N type during its post-maximum decline. \citet{Izz12} suggested the multiple absorptions seen in the P-Cygni profiles during the early phase to be due to the presence of a clumpy wind surrounding the white dwarf. Interferometric observations in the broadband and medium spectral resolution Br$\gamma$ line by \citet{Che11} indicated a bipolar event with an inclination angle of i = 15$^\circ$, and a position angle, PA = 110$^\circ$. Using high resolution optical spectroscopic observations, \citet{Sho13} modelled the ejecta using an axisymmetric conical, bipolar, geometry with an inclination angle of i = 15 $\pm$ 5$^\circ$, and estimated the ejected mass to be $M_{\rm{ej}}=2\, \times\, 10^{-6}\, M_\odot$. Using period change during the 2011 eruption, the ejected mass was estimated as $3\, \times\, 10^{-5}\, M_{\odot}$ or more by \citet{Pat17}, while \citet{Nel14} estimated the ejected mass as $(1-30)\, \times\, 10^{-5}\, M_{\odot}$ using the high peak flux densities in the radio emission. Based on the fact the system had a late turn-on time for the SSS phase, a large ejecta mass $\gtrapprox\, 10^{-5}\, M_{\odot}$ was suggested by \citet{cho14}. From a study of the IR photometric properties, \citet{Eva12} found a weak, cool IR excess that was attributed to the heating of a pre-outburst, dust in the swept up interstellar environment of the nova. T Pyx was detected as a supersoft source from day 105 to 349. The X-ray emission peak detected by {\it{Swift}} satellite was found to be compatible with the occurrence of high ionization lines in the optical spectrum, such as [Ne III], [C III] and N III (4640 \AA) around day 144. The occurrence of coronal lines like [Fe VII] and [Fe X] was found to be in line with the peak of radio emission around day 155, and the plateau phase in X-ray \citep{Sur14}. 

T Pyx is the only recurrent nova that has a discernible shell that was first detected by \citet{Due79}. Subsequent observations using the Hubble Space Telescope \citep{Sha89,Sha97,Sch10} revealed the shell to be expanding very slowly, and consisting of several knots. Based on three-dimensional, gas dynamical simulations of the evolution of the ejecta of T Pyx, \citet{Tor13} predicted the observed expansion of the shell and its morphology. Their simulations demonstrated that the knots are formed due to Richtmyer-Meshkov instabilities that set in when the ejecta from later outbursts collide with the older, swept-up, cold, dense shell.

We present in this paper the spectral evolution of T Pyx 2011 outburst and post outburst  based on spectra obtained from its time since discovery at t = 1.28 (pre-maximum phase) to 2415.62 days (late post-outburst phase). The evolution is studied through 1D and 3D photo-ionization modelling of the observed spectra at different phases. The results of linear polarimetric observation from day 1.36 to 29.33 in the BVRI bands are also discussed. 
\section{Observations} \label{2.1}
The optical spectroscopic and polarimetric observations were carried out at the following observatories:
\subsection{Spectroscopy}
\subsubsection{Indian Astronomical Observatory (IAO)}
Low resolution spectroscopic observations were obtained using the Himalayan Faint Object Spectrograph Camera (HFOSC) mounted on the 2 m Himalayan Chandra Telescope (HCT) located at IAO, Hanle, India. Spectra were obtained from 2011 April 15 (JD 2455667.08) to 2017 Nov 23 (JD 2458081.42) using grism 7 (wavelength range: 3500 -- 8000 \AA) with a resolution of R$\sim$1300 and grism 8 (wavelength range: 5200 -- 9200 \AA) with R$\sim$2200. The log of observations is given in \Cref{log}. Data reduction was performed in the standard manner using the various tasks in the IRAF\footnote{IRAF is distributed by the National Optical Astronomy Observatories, which are operated by the Association of Universities for Research in Astronomy, Inc., under cooperative agreement with the National Science Foundation.}. All the spectra were bias subtracted and extracted using the optimal extraction method. FeNe and FeAr arc lamp spectra were used for wavelength calibration. The instrumental response was corrected using spectrophotometric standard stars which were observed on the same night. For those nights, where the standard star spectra were not available, standard star spectra from nearby nights were used. The flux calibrated spectra in red and blue regions were combined and scaled to a weighted mean in order to obtain the final spectra. The zero points required to convert the spectra to the absolute flux scale were obtained from AAVSO magnitudes in UBVRI filters.
\begin{table}
	\centering
    \caption{Observational log for low resolution spectra}
    \label{log}
    \resizebox{\columnwidth}{!}{%
    \begin{tabular}{ccccc}
    	\hline
    	\hline
		\textbf{JD} & \textbf{Date} & \textbf{t (days)} & \textbf{Coverage (\AA)} & \textbf{Telescope}\\
 		\hline
		2455667.08 & 15 Apr 2011 & 1.28 & 3800 - 9200 & HCT, IGO\\
		2455668.08 & 16 Apr 2011 & 2.28 & 3800 - 9200 & HCT\\
    	2455670.23 & 18 Apr 2011 & 4.43 & 3900 - 8250 & IGO\\
		2455671.26 & 19 Apr 2011 & 5.46 & 3900 - 8250 & IGO\\
		2455672.14 & 20 Apr 2011 & 6.34 & 3800 - 9200 & HCT\\
        2455673.37 & 21 Apr 2011 & 7.57 & 3900 - 8250 & IGO\\
		2455674.29 & 22 Apr 2011 & 8.49 & 3900 - 8250 & IGO\\
		2455680.11 & 28 Apr 2011 & 14.31 & 3800 - 9200 & HCT\\
		2455681.09 & 29 Apr 2011 & 15.29 & 3800 - 9200 & HCT\\
		2455682.09 & 30 Apr 2011 & 16.29 & 3800 - 9200 & HCT\\
        2455688.09 & 06 May 2011 & 22.29 & 3800 - 8500 & VBT\\
		2455692.07 & 10 May 2011 & 26.27 & 3800 - 8800 & VBT\\
		2455694.07 & 12 May 2011 & 28.27 & 3800 - 8500 & VBT, IGO\\
		2455708.10 & 26 May 2011 & 42.30 & 3800 - 8500 & VBT\\
		2455720.09 & 07 Jun 2011 & 54.29 & 3800 - 8500 & VBT\\
		2455734.08 & 21 Jun 2011 & 68.28 & 3800 - 8500 & VBT\\
		2455889.46 & 23 Nov 2011 & 223.66 & 3800 - 6800 & HCT\\
		2455890.45 & 24 Nov 2011 & 224.65 & 3800 - 6800 & HCT\\
		2455896.42 & 30 Nov 2011 & 230.62 & 3800 - 9000 & HCT\\
		2455918.43 & 22 Dec 2011 & 252.63 & 3800 - 9000 & HCT\\
		2455932.38 & 05 Jan 2012 & 266.58 & 3800 - 8600 & HCT\\
        2455946.30 & 19 Jan 2012 & 280.50 & 4000 - 8600 & VBT\\
		2455947.36 & 20 Jan 2012 & 281.56 & 4000 - 8600 & HCT\\
		2455985.23 & 27 Feb 2012 & 319.43 & 4000 - 8600 & HCT\\
		2456002.20 & 15 Mar 2012 & 336.40 & 4000 - 8300 & HCT\\
        2456005.20 & 17 Mar 2012 & 339.40 & 4200 - 8300 & VBT\\
		2456031.08 & 13 Apr 2012 & 365.28 & 4000 - 9000 & HCT\\
		2456314.60 & 21 Jan 2013 & 648.80 & 4000 - 8600 & HCT\\
		2456730.19 & 13 Mar 2014 & 1064.39 & 4000 - 7500 & HCT\\
		2457387.34 & 30 Dec 2015 & 1721.54 & 4000 - 7500 & HCT\\
        2457419.24 & 31 Jan 2016 & 1753.44 & 4000 - 7500 & HCT\\
        2457809.17 & 24 Feb 2017 & 2143.37 & 4200 - 7500 & HCT\\
        2458081.42 & 23 Nov 2017 & 2415.62 & 4100 - 7500 & HCT\\

 		\hline
	\end{tabular}}
\end{table}
\subsubsection{IUCAA Girawali Observatory (IGO)}
Low resolution spectroscopic observations were also obtained using the IUCAA Faint Object Spectrometer and Camera (IFOSC), on the 2 m IGO telescope located at Girawali, India. Spectra were obtained from 2011 April 15 (JD 2455667.10) to 2011 May 13 (2455694.50) in the wavelength ranges of 3800 -- 6840 \AA\ and 5800 -- 8350 \AA\ using grisms IFOSC C7 with R$\sim$1100 and C8 with R$\sim$1600. HeNe lamp spectra were used for the wavelength calibration and flux calibration was carried out in the standard manner as above. The log of observations is given in \Cref{log}.

\subsubsection{Vainu Bappu Observatory (VBO)}
Low resolution spectroscopic observations were also obtained using the Opto Mechanics Research (OMR) spectrograph at 2.3 m Vainu Bappu Telescope (VBT) located at VBO, Kavalur, India. Spectra which cover a range of 3800 to 8800 \AA\ with R$\sim$1400 were obtained from 2011 May 6 (2455688.09) to 2012 March 17 (2456005.20). The log of observations is given in the \Cref{log}. Wavelength calibration was done using the FeNe and FeAr lamp spectra. Spectroscopic standards, HD93521 and Hiltner 600 observed on the same nights as the target were used to correct for the instrumental response.

High resolution spectra were obtained with the fiber-fed Echelle spectrograph at VBT. They cover a wavelength range from 4000 - 10000 \AA\ with resolution R = 27000. Reduction of these spectra was carried out using standard tasks in IRAF like bad pixel removal, scattered light subtraction, bias corrections, flat fielding and aperture extraction. ThAr lamp was used for the wavelength calibration. The log of observations is given in the \Cref{log2}.
\begin{table}
	\centering
    \caption{Observational log for high resolution spectra}
    \label{log2}
    \resizebox{0.55\columnwidth}{!}{%
    \begin{tabular}{ccc}
    	\hline
    	\hline
		\textbf{JD} & \textbf{Date} & \textbf{t (days)}\\
 		\hline
		2455668.10 & 16 Apr 2011 & 2.31 \\
        2455669.14 & 17 Apr 2011 & 3.35 \\ 
        2455678.11 & 26 Apr 2011 & 12.32 \\ 
        2455695.08 & 13 May 2011 & 29.29 \\
        2455696.07 & 14 May 2011 & 30.28 \\
        2455704.07 & 22 May 2011 & 38.28 \\ 
        2455722.08 & 09 Jun 2011 & 56.29 \\ 
        2455723.08 & 10 Jun 2011 & 57.29 \\ 
 		\hline
	\end{tabular}}
\end{table}
\subsection{Polarimetry}
Linear polarization data were obtained using IFOSC in imaging polarimetry mode in $B$, $V$, $R$, and $I$ filters covering the wavelength range of 3500 -- 8000 \AA. The design is a standard one with stepped half-wave plate followed by a Wollaston prism; a focal mask is used to prevent the ordinary and extra ordinary images overlapping \citep{ram98}. Polarimetric observations were obtained from day 1.36 to 29.33 during the initial phase for eight nights. Polarized standard star, HD160529 was observed on day 2.71 and HD147084 was observed from day 2.71 to 7.71. Unpolarized standard star, HD98281 was observed from day 2.71 to 7.71. The instrumental polarization correction obtained from unpolarized standard star was estimated to be 0.1\%, and position angle correction was estimated from polarized standard stars. The observed degree of polarization and position angle for the polarized standard stars are listed in \Cref{pol_std}. Observations were carried out at four different positions (0, 22.5, 45, 67.5) of the half wave plate. Data were analysed using IRAF. Pre-processing such as bias subtraction was carried out on all the frames. Ordinary and extraordinary image pairs were identified, and aperture photometry was performed on all the frames. The ratio of counts in ordinary and extraordinary images was used to estimate normalized $q$ and $u$. Interstellar polarization correction has not been applied here.
\begin{table}
\caption{Polarimetric observations of polarized standard stars}
\label{pol_std}
\begin{center}
\resizebox{0.95\columnwidth}{!}{%
\begin{tabular}{cccccc}
\hline
\textbf{Object}   & \textbf{JD }        & \textbf{Filter} & \textbf{P $(\%)$}  & \textbf{$\theta$} \\
\hline
\hline 
HD160529 & 2455669.17 & B      & 7.27$\pm$0.09 & 66.82$\pm$0.34  \\ 
         &            & V      & 7.49$\pm$0.05  & 66.48$\pm$0.18  \\
         &            & R      & 7.14$\pm$0.07 & 64.82$\pm$0.30  \\
         &            & I      & 6.01$\pm$0.06 & 65.87$\pm$0.28  \\
         \cdashline{3-5}
         & & B$^{1}$ & 7.46$\pm$0.04 & 20.1 \\
         & & V & 7.76$\pm$0.03 & 20.4  \\
         &            & R     & 7.41$\pm$0.06    & 21.7   \\
         &            & I      & 5.49$\pm$0.03    & 21.1  \\
         \cdashline{3-5}
         &            & B$^{2}$      & 7.24         & 20.1  \\
         &            & V      & 7.52         & 20.1 \\
         \cline{1-5}
HD147084 & 2455669.17 & B      & 3.48$\pm$0.09    & 80.04$\pm$0.75 \\
         &            & V      & 4.18$\pm$0.05    & 79.04$\pm$0.45   \\
         & 2455671.11 & B      & 3.34$\pm$0.07    & 81.63$\pm$0.61   \\
         &            & V      & 3.98$\pm$0.05    & 78.75$\pm$0.40  \\
         & 2455673.21 & B      & 3.48$\pm$0.05    & 80.32$\pm$0.45   \\
         &            & V      & 4.15$\pm$0.06    & 79.72$\pm$0.45 \\
         & 2455674.12 & B      & 3.51$\pm$0.16    & 55.14$\pm$1.90  \\
         &            & V      & 4.08$\pm$0.05    & 55.73$\pm$0.32 \\
         \cdashline{3-5}
         &            & B$^{3}$      & 3.5          & 32  \\
         &            & V      & 4.18         & 32  \\
         \cdashline{3-5}
         &            & V$^{4}$       & 4.12$\pm$0.02    & 32.2 \\
         &            &        & 4.16$\pm$0.01    & 32.10$\pm$1.90   \\
         \cline{1-5}
\end{tabular}}
\end{center}
$^{1}$\citet{Gos13}; $^{2}$\citet{Cla98};\\$^{3}$ \citet{ser74} and $^{4}$\citet{Gos10}
\end{table}

\section{Analysis of the observed data}
\subsection{Evolution of the spectra} \label{2.2}
The evolution of the optical spectrum from the pre-maximum to the post-outburst phase is described here. The date of outburst discovery i.e. 2011 April 14.29 UT (JD 2455665.79) is considered as t=0.

\begin{figure*}
\includegraphics[scale=0.4]{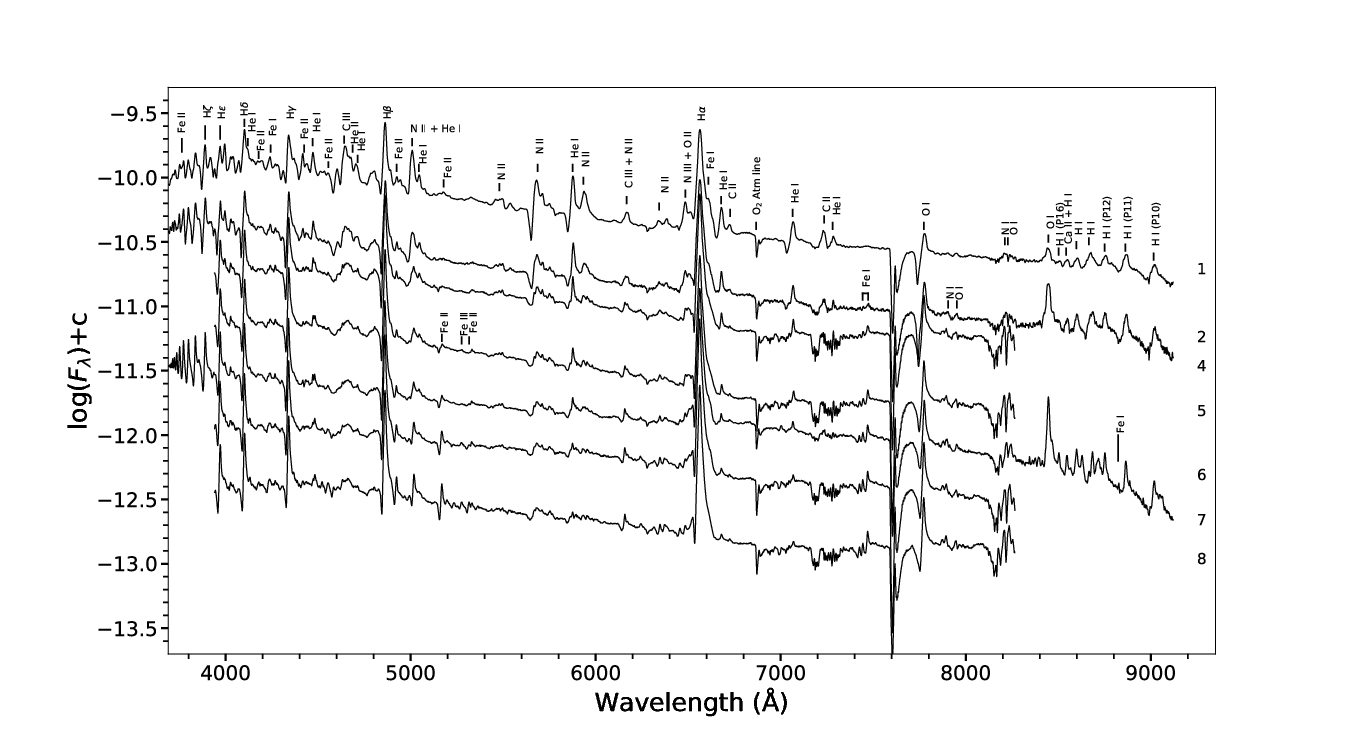}
\caption{Low-resolution optical spectral evolution of optically thick phase of T Pyx from 2011 April 15 to 2011 April 22. The spectra are dominated by P-Cygni profiles. The identified lines and time since discovery in days (numbers to the right) are marked. The P-Cygni absorption components become sharper and deeper as the system evolves.}
\label{phase1}
\end{figure*}

\begin{figure*}
\centering
\includegraphics[scale=0.4]{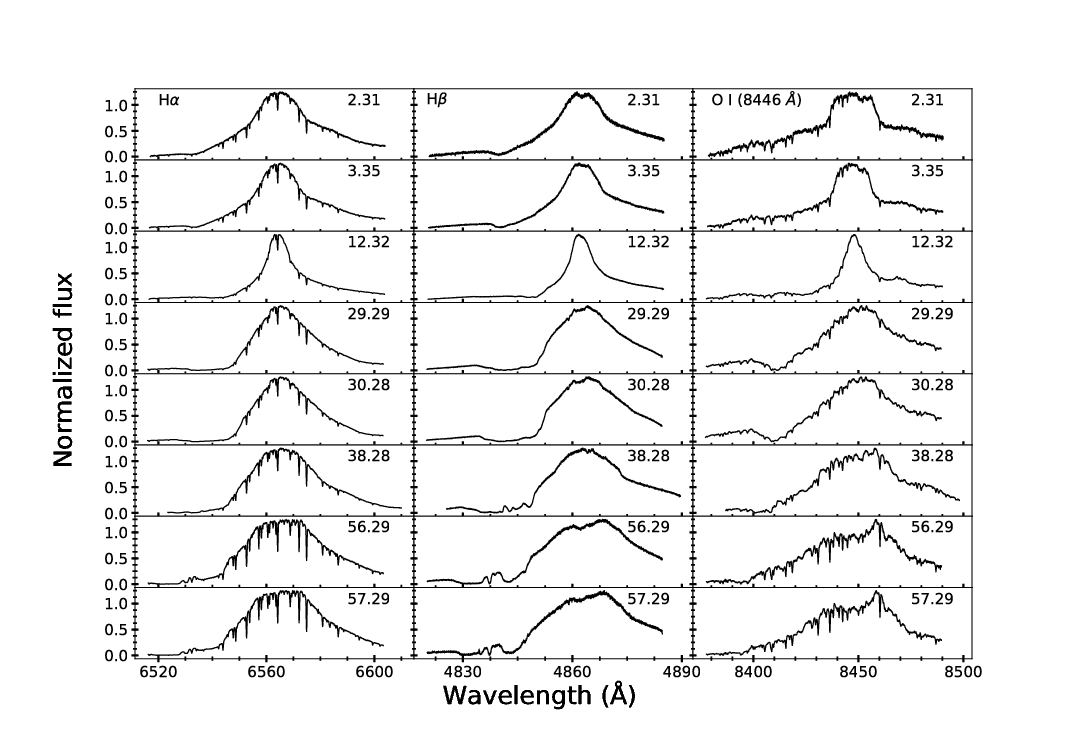}
\caption{High resolution spectral (R = 27000) evolution of T Pyx from 2011 April 16 to 2011 June 10. Evolution of H$\alpha$, H$\beta$ and O I (8446 \AA) line profiles are shown. Notice the variation in the velocity of the lines as the system evolves. There is decrease in velocity from day 2.31 to 12.32 and then increase up to $\sim$2000 km s$^{-1}$ till day 57.29.}
\label{hho}
\end{figure*}

Spectra from days 1.28 to 8.49 i.e., the initial optically thick phase before optical maximum, shown in \Cref{phase1} consist of strong hydrogen Balmer, oxygen, helium, neon, carbon, calcium, nitrogen and a few iron lines. The strength of hydrogen Balmer, Fe II, O I and N I lines increase with time, while high excitation lines like He I, Ne I, Ne II, O II + N III, and C III drop in intensity and some lines eventually disappear at the end of this phase. Iron lines become prominent towards the end of this phase. A decrease in the ejecta velocity, from $\sim$ 2500 km s$^{-1}$ to $\sim$ 1000 km s$^{-1}$, was observed until day 12.32 (\Cref{hho}). On day 1.28, hydrogen Balmer, O I (7774 \AA) and N II lines have P-Cygni profile with blue-shifted absorption components. These lines develop deeper and sharper P-Cygni absorption components and narrower emission components and slowly fades away. The other elements which have P-Cygni profiles are Fe II and He I. This phase marks the end of the fireball stage (pseudo-photospheric expansion).

\begin{figure*}
\centering
\includegraphics[scale=0.4]{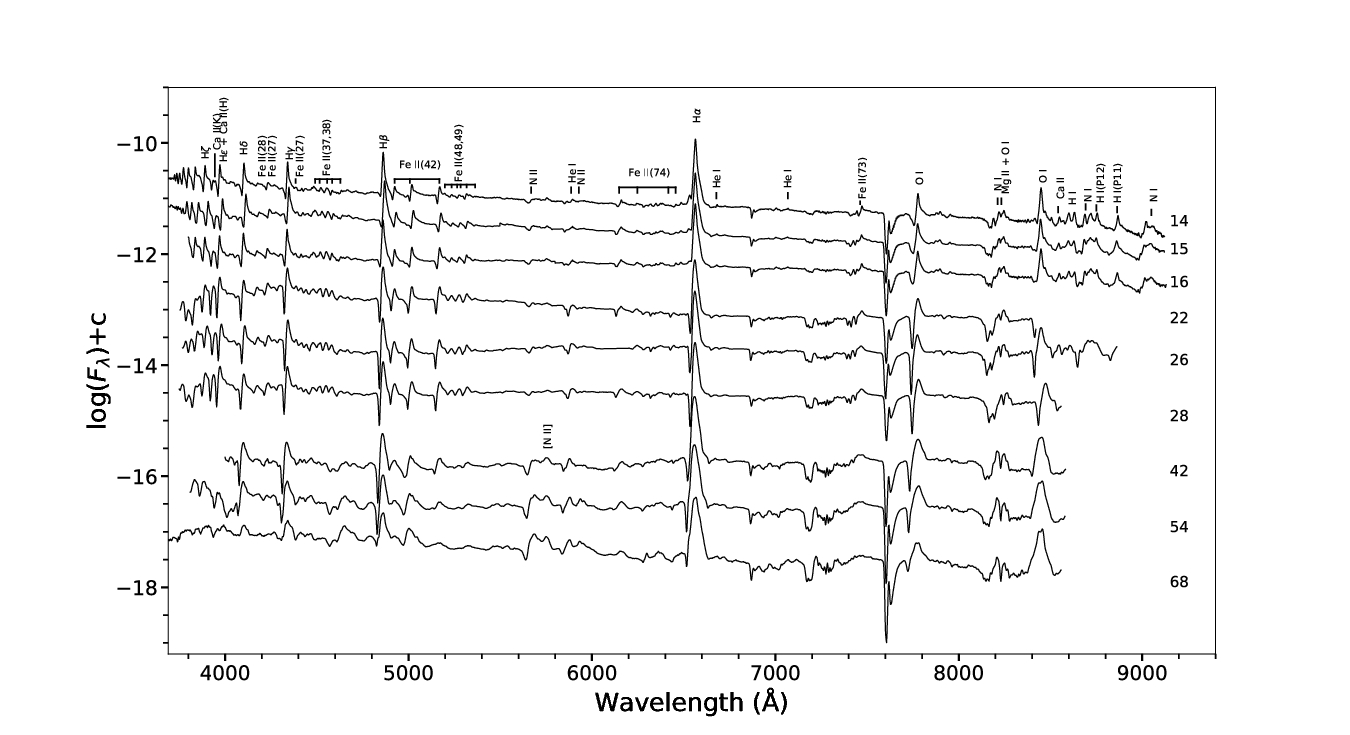}
\caption{Evolution of the spectra as the T Pyx system passes through its optical maximum and early decline phase from 2011 April 28 to 2011 June 21. The significant development in this phase is the presence and slow evolution of Fe II multiplet profiles from P-Cygni to emission lines. The identified lines and time since discovery in days (numbers to the right) are marked.}
\label{phase2}
\end{figure*}

Fe II multiplets (\Cref{phase2}) are the most dominant non-Balmer lines as the system evolves to its optical maximum, and during the early decline (days 14.31 to 68.28). O I (8446 \AA) line which was an emission line till the previous phase develops a P-Cygni profile from day 22.29 to 28.27 (around the optical maximum). Fe II multiplets, Balmer and O I lines which have P-Cygni profiles slowly evolve into emission lines towards the end of this phase. H$\alpha$ line becomes broader with a velocity up to $\sim$2000 km s$^{-1}$ and round peaked near the optical maximum. FWHM of the lines increases from $\sim$1000 to $\sim$2000 km s$^{-1}$ during day 14.31 to 68.28. From day 42.30 to 68.28, presence of [N II] (5755 \AA) and N II lines are clearly seen. The emission component of N II (5679 \AA) P-Cygni profile which is weak on $\sim$day 14 becomes stronger by $\sim$day 42.

\begin{figure*}
\includegraphics[scale=0.4]{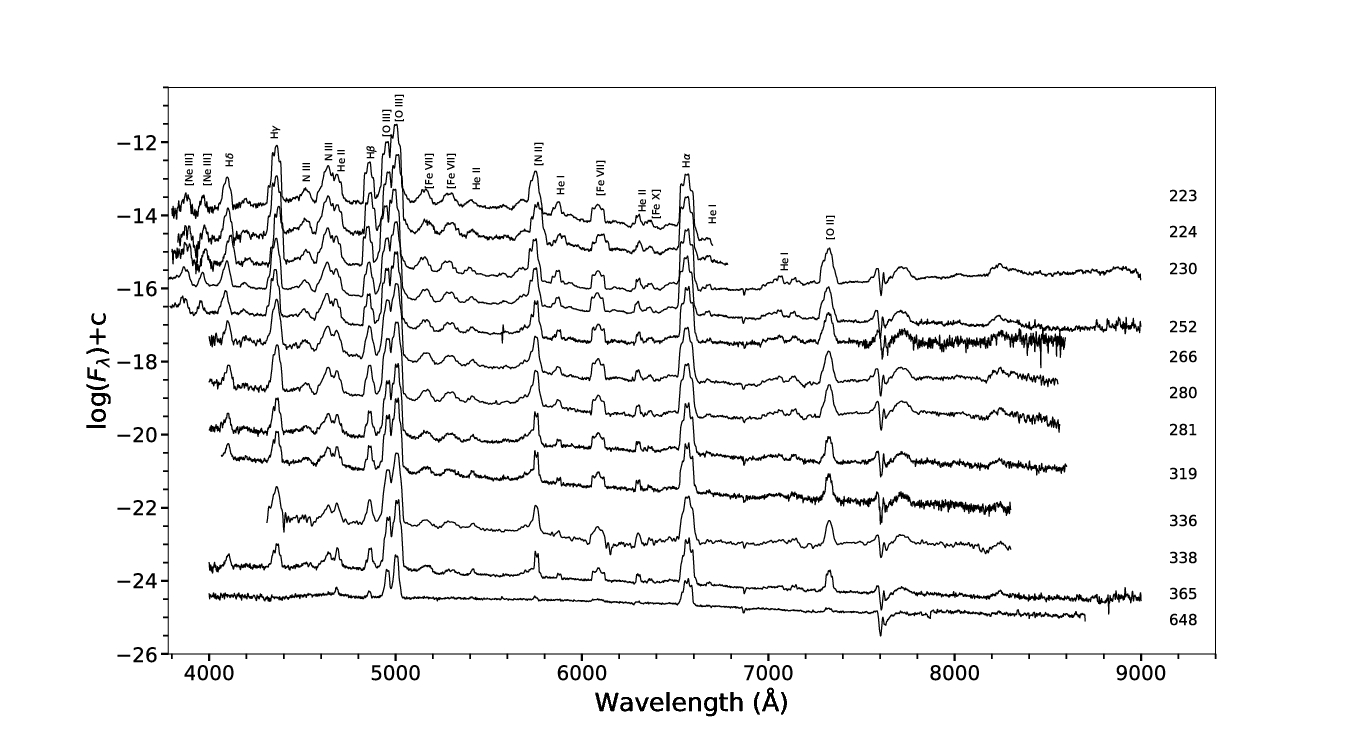}
\caption{Evolution of the spectra during nebular phase of T Pyx from 2011 Nov 23 to 2013 Jan 21. The emission line profiles are double-peaked with wing-like structures on both the sides. Presence of the forbidden lines can clearly be seen during this phase. The identified lines and time since discovery in days (numbers to the right) are marked.}
\label{nebu}
\end{figure*}

\begin{figure*}
\includegraphics[scale=0.4]{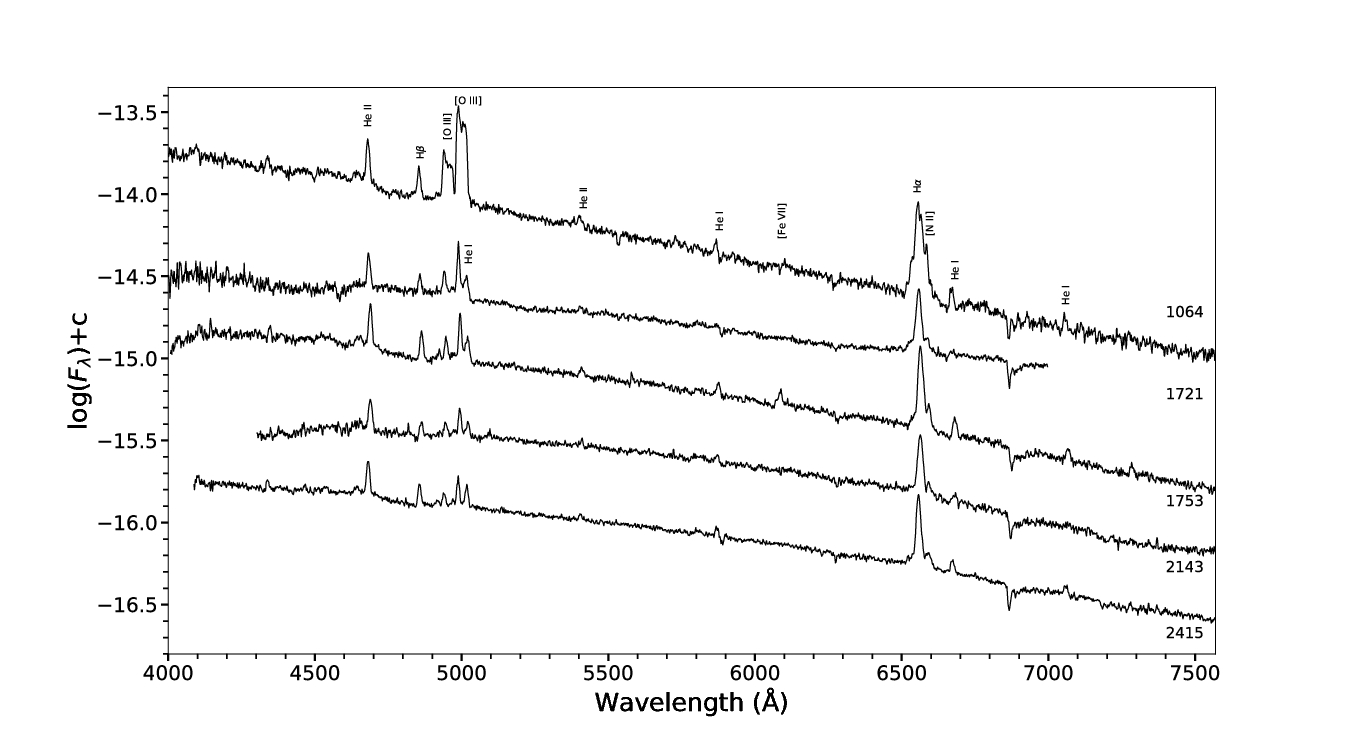}
\caption{Evolution of the spectra during the late post-outburst phase of T Pyx from 2014 Mar 13 to 2017 Nov 23. As the system evolves, the lines develop into narrow emission ones from broader profiles. The [O III] lines are blue-shifted by $\sim$780 km s$^{-1}$ beyond day 1721 and [N II] 6874 \AA\ line can clearly be seen in this phase. The identified lines and time since discovery in days (numbers to the right) are marked.}
\label{quie}
\end{figure*}

The nebular phase (day 223.66 to 648.80) is marked by dominant, broad [O III] 4959 and 5007 ~\AA\ lines (\Cref{nebu}). The [N II] line develops double-peaked profile with wing-like structures on either side which is similar to the [O III] and the Balmer lines. Other forbidden lines seen are [Fe VII] (5158, 5276 and 6087 \AA), [Fe X] (6375 \AA), [O II] (7330 \AA), [C III]+[O III] (Blend 4364 \AA\ with H I) and [Ne III] (3869 and 3968 \AA) lines. The intensity of [N II] and [Fe VII] lines slowly decreases and [N II] disappears by day 1064. By day 1064, all the nebular lines disappear, except for the [O III] 4959, 5007 ~\AA\ lines. The intensity of [O III] lines reduced during the nova's decline to late post-outburst phase during day 1064.39 to 2415.62 (\Cref{quie}). 

Strong helium emission lines are seen during this phase. The [O III] lines show a blueshift by $\sim$ 780 km s$^{-1}$ beyond day 1721, which is not seen in the hydrogen and helium emission lines. This could be due the fact that while the [O III] lines arise in the fading nova ejecta, the hydrogen and helium lines arise in the accretion disc. 

\subsection{Polarization of the ejecta}\label{3.2}
The degree of polarization and position angle values estimated for T Pyx during its 2011 outburst are given in \Cref{pol_tpy}. During the initial rise phase, the degree of polarization is found to increase, in all bands, until day 4-5, and decrease subsequently. For example, the degree of polarization in the $V$ increased from a value of 0.69\% on day 1.36 to 1.27\% on day 5.35, and decreased to 0.37\% by day 8.34. The degree of polarization was found to have increased from the estimate on day 8.34, during the next set of observations on days 28.34 and 29.33, with a rising trend. The position angle was found to be 112$^\circ$ $\pm$ 18$^\circ$ during the entire period of observation.
\begin{table}
\caption{Polarimetric observations of T Pyx during its 2011 outburst}
\label{pol_tpy}
\begin{center}
\resizebox{0.8\columnwidth}{!}{%
\begin{tabular}{ccccc}
\hline
\textbf{JD} & t (days) & \textbf{Filter} & \textbf{P $(\%)$} & \textbf{$\theta$} \\
\hline
\hline
2455667.14 & 1.36 & B & 0.650$\pm$0.063 & 102.05$\pm$2.57 \\
            & & V & 0.695$\pm$0.062 & 104.35$\pm$4.09 \\
          & & R & 0.727$\pm$0.083 & 112.92$\pm$3.28 \\
          & & I & 0.819$\pm$0.191 & 105.06$\pm$5.64 \\
           \cdashline{1-5}
 2455669.2 & 3.39 & V & 0.849$\pm$0.073 & 106.71$\pm$2.43 \\
           & & R & 0.858$\pm$0.081 & 104.11$\pm$2.64 \\
           & & I & 0.638$\pm$0.104 & 108.05$\pm$3.20 \\ 
            \cdashline{1-5}
 2455670.12 & 4.33 & B & 0.785$\pm$0.028 & 94.98$\pm$2.36 \\
           & & V & 0.844$\pm$0.048 & 93.47$\pm$2.06 \\
            & & R & 0.903$\pm$0.076 & 105.29$\pm$2.23 \\
         & & I & 0.978$\pm$0.062 & 101.09$\pm$1.84 \\ \cdashline{1-5}
2455671.14 & 5.35 & B & 1.250$\pm$0.062 & 100.18$\pm$1.42 \\
          & & V & 1.270$\pm$0.062 & 102.14$\pm$1.41 \\
           & & R & 1.132$\pm$0.063 & 102.48$\pm$1.59 \\
          & & I & 0.928$\pm$0.082 & 104.03$\pm$2.48 \\ \cdashline{1-5}
 2455672.12 & 6.33 & B & 0.590$\pm$0.028 & 101.99$\pm$2.88 \\
           & & V & 0.492$\pm$0.052 & 113.30$\pm$2.99 \\
          & & R & 0.575$\pm$0.046 & 115.12$\pm$2.41 \\
           & & I & 0.335$\pm$0.068 & 112.82$\pm$5.81 \\ \cdashline{1-5}
2455673.21 & 7.42 & B & 0.235$\pm$0.037 & 129.40$\pm$4.73 \\ \cdashline{1-5}
2455674.13 & 8.34 & B & 0.383$\pm$0.037 & 113.41$\pm$1.23 \\
            & & V & 0.369$\pm$0.051 & 129.02$\pm$2.51 \\
            & & R & 0.322$\pm$0.026 & 100.97$\pm$4.11 \\
            & & I & 0.352$\pm$0.051 & 128.75$\pm$4.01 \\ \cdashline{1-5}
2455694.15 & 28.34 & B & 0.549$\pm$0.029 & 118.51$\pm$5.56 \\
         & & V & 0.538$\pm$0.037 & 112.04$\pm$1.66 \\     
         & & R & 0.663$\pm$0.032 & 116.28$\pm$1.62 \\
            & & I & 0.515$\pm$0.051 & 122.81$\pm$2.94 \\ \cdashline{1-5}
 2455695.11 & 29.33 & B & 0.647$\pm$0.059 & 127.51$\pm$6.44 \\
           & & V & 0.741$\pm$0.055 & 121.30$\pm$1.90 \\
           & & R & 0.875$\pm$0.080 & 118.41$\pm$3.08 \\
           & & I & 0.750$\pm$0.049 & 122.19$\pm$1.84 \\
\hline
\end{tabular}}
\end{center}
\end{table}

\begin{figure}
\centering
  \resizebox{\hsize}{!}{\includegraphics{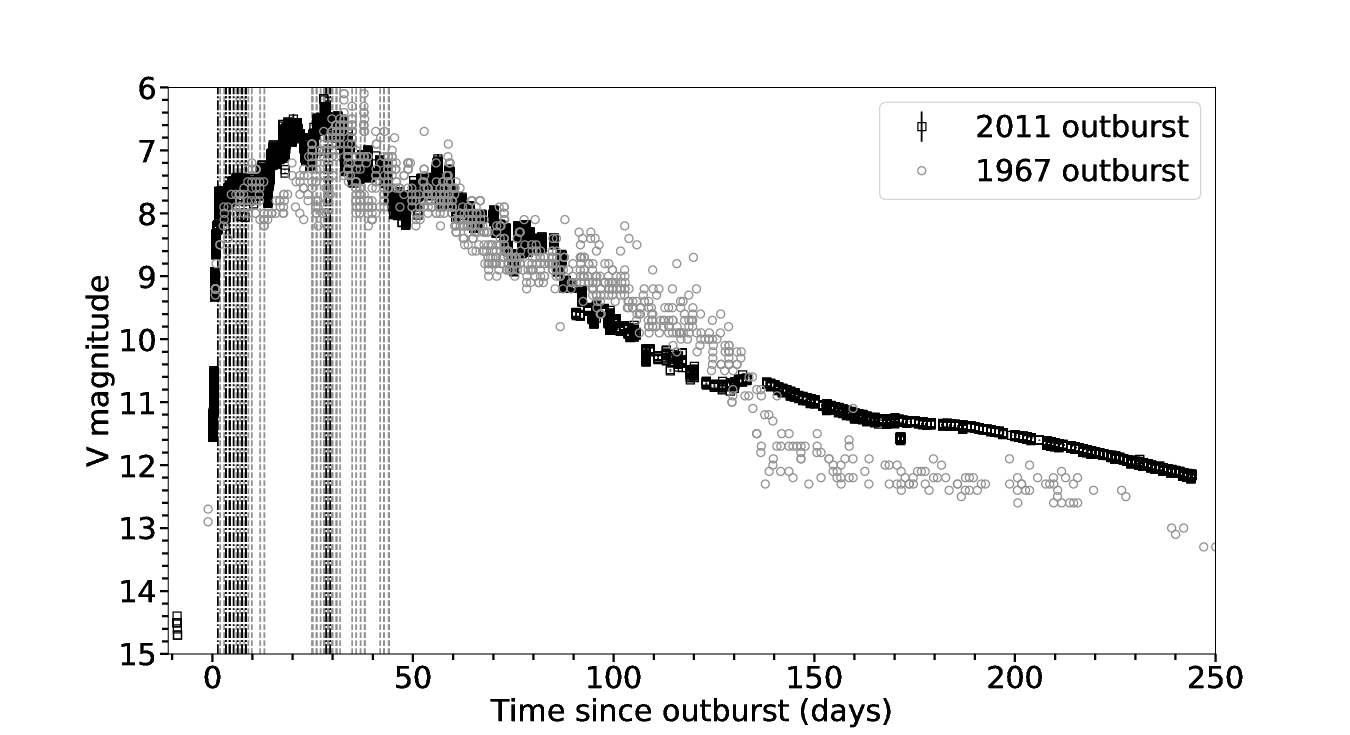}}
  \caption{The light curve for T Pyx during its 1967 and 2011 eruptions. The black squares  and gray circles correspond to the 2011 and 1967 eruption respectively. The black and gray dash lines correspond to the polarization epochs during 2011 and 1967 eruption respectively.}
  \label{lightcurve}
\end{figure}

\begin{figure}
  \resizebox{\hsize}{!}{\includegraphics{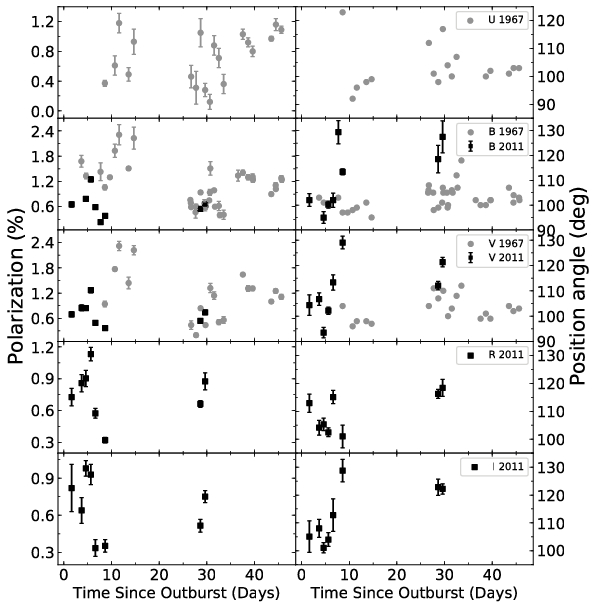}}
  \caption{The variability in the degree of polarization and position angle for 2011 and 1967 outbursts. The squares and circles correspond to 2011 and 1967 outburst respectively.}
  \label{polplot}
\end{figure}

Intrinsic polarization of the ejecta was detected by \cite{Egg67} during the 1967 outburst, that was found to vary with time. Polarization values of T Pyx during its 2011 outburst has been compared with those reported for the 1967 outburst during days 1.96--44.02. In \Cref{lightcurve}, the epochs of polarization observations made during the two outbursts are highlighted. The variability in degree of polarization values in different filters for both the outbursts is shown in \Cref{polplot}. An identical pattern in the variation of degree of polarization is observed in both the outbursts i.e., an initial rise in the values followed by a decrease in the early pre-maximum phase, followed by an increase in the degree of polarization during the optical maximum. While the trend is similar, it is noticed that the degree of polarization is lower in the 2011 outburst compared with the 1967 outburst. The maximum value of polarization in B filter and V filter are $2.3\%$ and $2.32\%$ respectively for 1967 outburst, while for 2011 outburst it is $1.25\%$ and $1.27\%$ in B and V filters respectively. Position angles observed during 2011 outburst are consistent with those of the 1967 outburst (\Cref{polplot}). Although both the 1967 and 2011 values are uncorrected for interstellar polarization, the very similar behaviour during both epochs indicates intrinsic polarization.

\section{Photo-ionization analysis}\label{sec4}
The evolution of the emission spectrum during and after the 2011 outburst was modelled to understand the physical conditions, and the variation in the physical conditions as the system evolves. The photo-ionization code CLOUDY, C17.00 \citep{Fer17} was used to model the evolution. CLOUDY has been applied to determine the physical characteristics and elemental abundances of novae in a few previous studies \citep{Sch01,Sch02,Van05,Sch07a,Sch07b,Hel10,Das15,Mon18,Raj18}. Spectra obtained at epochs of day 68, 224, 252, 336 and 1064 corresponding to different phases were modelled. The modelled synthetic spectrum was compared with the observed spectrum and parameters such as elemental abundances, density, source luminosity, source temperature etc. were obtained. These parameters were used as inputs to model the 3D ionization structure of the ejecta. 

To obtain the 3D ionization structure of the ejecta, pyCloudy \citep{Mor13} was used. Modelling the 3D ionization structure of the ejecta would reveal spatial distribution of elements at different epochs and also their evolution with time. This would result in understanding the origin of different lines at different phases. This work is an attempt to understand the spatial origin of the ionization lines which is not well understood. The pyCloudy is a pseudo-3D code with CLOUDY 1D \citep{Fer17} photo-ionization code and python library to analyse the models. The 3D CLOUDY model can be obtained by generating 1D CLOUDY fits corresponding to different angles and interpolating these runs (radial radiation) to form a 3D coordinate cube. Python library includes the routines in order to visualize the models, to integrate masses, fluxes, etc. The photo-ionization models used in pyCloudy are 1D codes and they do not deal with the effect of diffuse radiation field entirely, also Python codes are limited by symmetry, hence it is referred to as pseudo-3D and not complete 3D code. Using photo-ionization modelling is usually a good enough approximation to study physical conditions in novae. Full 3D codes which use Monte Carlo methods and hydrodynamic calculations will be considered at a later point in time.

Many parameters are used in CLOUDY to define the initial physical conditions of the source and the ejected shell. These parameters include shape and intensity of the external radiation field striking a cloud, chemical composition of gas, and geometry of the gas, including its radial extent and dependence of density on radius. CLOUDY generates predicted spectrum using these input parameters by solving the equations of thermal and statistical equilibrium from non-LTE, illuminated gas clouds. The density, radii, geometry, distance, covering factor, filling factor and elemental abundances define physical conditions of the shell. The density of the shell is defined by hydrogen density. The radial variation of hydrogen density, n(r) and filling factor, f(r) are as follows:
\begin{align}
\label{eqn1}
\begin{split}
n(r) = n(r_{0})(\dfrac{r}{r_{0}})^{\alpha}\\
f(r) = f(r_{0})(\dfrac{r}{r_{0}})^{\beta}
\end{split}
\end{align}
where r$_{0}$ is the inner radius, $\alpha$ and $\beta$ are exponents of power laws. Here, $\alpha$ = -3  for a shell undergoing ballistic expansion, and the filling factor power law exponent, $\beta$ = 0 similar to previous studies \citep{Sch02,Van05,Hel10,Das15}. 

In the 1D model, the geometry of the shell was assumed to be a spherically symmetric, expanding one illuminated by the central source. Many spectra were generated to obtain the best-fit for each epoch by varying free parameters like hydrogen density, effective blackbody temperature and abundances of the elements which were seen in the observed spectrum, while the remaining elements were fixed at the solar values. The ejecta was assumed to be made of more than one density region in order to fit all the ionized lines. A covering factor was set in all the regions such that the sum was always equal to 1. All parameters, except for hydrogen density and the covering factor, were kept constant in all the regions, so that the number of free parameters from the regions were reduced. Modelled line ratios were obtained by adding line ratios of each region after multiplying by its covering factor. The inner and outer radii of ejected shell were set by the time of observation, and minimum and maximum expansion velocities obtained using full width at half maximum of all the emission lines, similar to the study by \cite{Sch02,Sch07b,Hel10,Mon18}. 

The best fit model was obtained by calculating $\chi^{2}$ and reduced $\chi^{2}$:
\begin{align}
\begin{split}
\label{eqn2}
\chi^{2} = \sum_{i=1}^{n} \dfrac{(M_{i} - O_{i})^{2}} {\sigma^{2}_{i}}\\
\chi^{2}_{red} = \dfrac{\chi^{2}}{\nu}
\end{split}
\end{align}
M$_{i}$ \& O$_{i}$ -- modelled \& observed line ratios,\\
$\sigma_{i}$ is the error in observed flux ratio\\
n -- number of observed lines\\
n$_{p}$ -- number of free parameters\\
$\nu$ = n - n$_{p}$, degrees of freedom.\\
The ejected mass was determined using the relation (e.g., \citealt{Sch01,Sch02,Das15}):
\begin{align}
\label{eqn3}
M_{shell} = n(r_{0})f(r_{0}) \int_{R_{in}}^{R_{out}} (\dfrac{r}{r_{0}})^{\alpha+\beta} 4 \pi r^{2} dr
\end{align}

The ejected mass was calculated for all the regions, then multiplied by the corresponding covering factors and added to obtain the final value. The abundance values and other parameters are obtained from the model. The abundance solutions are sensitive to changing opacity and physical conditions of the ejecta like temperature.  

The initial structure of the cloud for 3D modelling was assumed to be an axisymmetric spheroid. The input parameters for this central source were defined using the best-fit 1D CLOUDY results obtained for each epoch. Parameters adopted were effective blackbody temperature, abundance values, density, filling factor, inner and outer radii. Six 1D CLOUDY runs at different radial directions were used to build this 3D cube. All the 1D models with their respective radial direction are inclined to the equatorial plane at a specified angle with the inner and outer radii. In every 1D run, the elemental abundances and luminosity were kept constant while temperature, inner \& outer radii and density were varied. In every radial direction (1D run), the emissivity values of every element present in the observed spectrum were interpolated to obtain the 3D emissivity values for every element. The dimension of the 3D cube is 301 $\times$ 301 $\times$ 301.

\subsection{The early decline phase}
For this phase, the spectrum of day 68 was modelled. The central ionizing source was set to be at effective temperature, 10$^{5}$ K and luminosity 10$^{37}$ erg s$^{-1}$. Three different regions were used to obtain the synthetic spectrum. A diffused region with density 4.46 $\times$ 10$^{8}$ cm$^{-3}$ to fit most of the lines like helium, C III and [N II]. Fe II lines were fit dominantly by clumpy region with a higher density, 5.62 $\times$ 10$^{8}$ cm$^{-3}$. N II recombination lines were fit using a region with lower temperature, 3 $\times$ 10$^{4}$ K and a higher density, 5.62 $\times$ 10$^{8}$ cm$^{-3}$. 

\begin{figure*}[!b]
\includegraphics[scale=0.5]{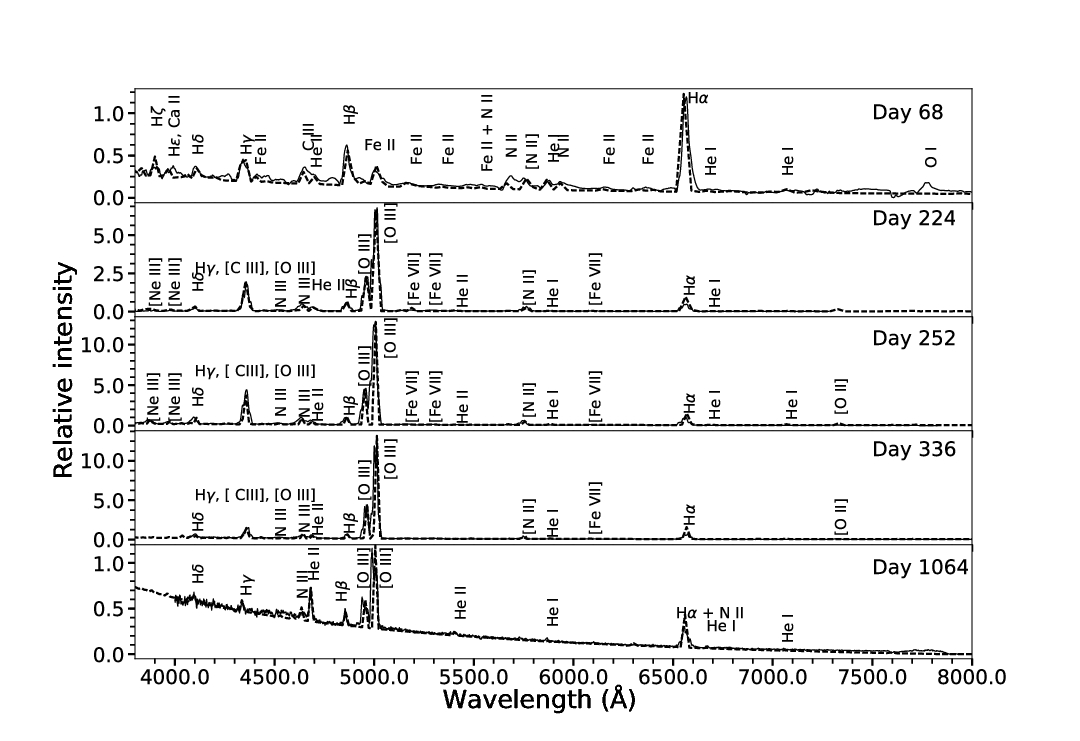}
\caption{Best-fit CLOUDY model spectra (dash line) plotted over the observed spectra (continuous line) of T Pyx obtained on day 68, 224, 252, 336 and 1064. The spectra are normalized to H$\beta$. The identified lines and time since discovery in days (numbers to the right) are marked.}
\label{ccc}
\end{figure*}

The relative fluxes of the observed lines, best-fit modelled lines and corresponding $\chi^{2}$ values are given in \Cref{cval_d68}, and the values of best-fit parameters are given in \Cref{param_d68}. The estimated abundance values show that nitrogen and helium abundances are more than solar, while iron, calcium and carbon abundance values are solar. The best-fit modelled spectrum (dash line) with the corresponding observed optical spectrum (continuous line) are as shown in \Cref{ccc}. Absorption components of the P-Cygni profiles are not modelled here because of the limitation of the code. The best-fit parameters obtained for this epoch is accurate to 65-70\% only.
\begin{table}
\caption{Observed and best-fit CLOUDY model line flux values$^{a}$ for day 68}
\label{cval_d68}
\begin{center}
 \resizebox{0.85\columnwidth}{!}{%
\begin{tabular}{l c c c c} 
 \hline
  \textbf{Line ID} & \boldmath{$\lambda$ (\AA)} & \textbf{Observed} & \textbf{modelled} & \boldmath{$\chi^{2}$}\\ 
 \hline\hline
 H I & 3889 & 2.30E$-$01 & 4.26E$-$01 & 9.45E$-$01 \\ [0.25ex]
Ca II (K) & 3934 & 1.18E$-$01 & 2.28E$-$01 & 1.60E$-$01 \\ [0.25ex]
H I, Ca II (H) & 3970 & 1.26E$-$01 & 2.40E$-$01 & 1.30E$-$01\\ [0.25ex]
H I & 4102 & 4.11E$-$01 & 3.09E$-$01 & 3.43E$-$01  \\ [0.25ex]
Fe II & 4178 & 2.82E$-$02 & 6.22E$-$02 & 1.64E$-$01 \\ [0.25ex]
Fe II & 4233 & 2.42E$-$02 & 4.85E$-$02 & 5.83E$-$03 \\ [0.25ex]
H I & 4340 & 1.11E$+$00 & 9.07E$-$01 & 4.71E$-$01 \\ [0.25ex]
Fe II & 4491 & 1.59E$-$01 & 3.96E$-$02 & 1.82E$-$01 \\ [0.25ex]
C III & 4650 & 4.48E$-$01 & 3.18E$-$01 & 6.98E$-$01 \\ [0.25ex]
He II & 4686 & 2.76E$-$01 & 1.86E$-$01 & 1.34E$-$01 \\ [0.25ex]
H I & 4861 & 1.00E$+$00 & 1.00E$+$00 & 0.00E$+$00 \\ [0.25ex]
Fe II & 4924 & 1.29E$-$01 & 6.58E$-$02 & 2.8E$-$01 \\ [0.25ex]
Fe II & 5018 & 7.32E$-$01 & 5.91E$-$01 & 2.98E$-$01 \\ [0.25ex]
Fe II & 5168 & 1.97E$-$01 & 1.33E$-$01 & 1.52E$-$01 \\ [0.25ex]
Fe II & 5235 & 5.40E$-$02 & 1.51E$-$01 & 1.41E$-$01 \\ [0.25ex]
Fe II & 5317 & 9.60E$-$02 & 3.78E$-$02 & 6.07E$-$02 \\ [0.25ex]
Fe II + N II & 5535 & 3.51E$-$02 & 9.27E$-$02 & 1.42E$-$01 \\ [0.25ex]
N II & 5679 & 8.02E$-$01 & 1.79E$-$01 & 8.81E$+$00 \\ [0.25ex]
[N II] & 5755 & 1.73E$+$00 & 2.90E$-$01 & 1.46E$-$01 \\ [0.25ex]
He I & 5876 & 1.76E$+$00 & 2.86E$-$01 & 1.68E$-$01 \\ [0.25ex]
N II & 5938 & 1.37E$-$01 & 4.89E$-$02 & 2.87E$-$01 \\ [0.25ex]
Fe II & 6148 & 1.02E$-$01 & 5.36E$-$02 & 5.11E$-$02 \\ [0.25ex]
Fe II + N II & 6248 & 3.42E$-$02 & 3.95E$-$02 & 8.93E$-$04 \\ [0.25ex]
Fe II & 6417 & 1.58E$-$01 & 5.50E$-$02 & 1.58E$-$01 \\ [0.25ex]
H I & 6563 & 3.54E$+$00 & 2.65E$+$00 & 9.03E$+$00 \\ [0.25ex]
He I & 6678 & 2.92E$-$02 & 5.19E$-$02 & 9.80E$-$03 \\ [0.25ex]
He I & 7065 & 2.99E$-$02 & 1.21E$-$01 & 1.10E$-$01 \\ [0.25ex]
\hline
\end{tabular}}
\end{center}
$^{a}$ Relative to H$\beta$ \\
$^{b}$ Calculated using \Cref{eqn2}
\end{table}

\begin{table}
\caption{Best-fit CLOUDY Model parameters for day 68}
\label{param_d68}
\begin{center}
\resizebox{0.8\columnwidth}{!}{%
\begin{tabular}{l c} 
 \hline
  \textbf{Parameter} & \textbf{Day 68} \\ [0.5ex] 
 \hline\hline 
 T$_{BB}$ ($\times$ 10$^{5}$ K) & 1.00\\ [0.25ex]
Luminosity ($\times$ 10$^{37}$ erg/s) & 1.00\\ [0.25ex]
Clump Hydrogen density ($\times$ 10$^{8}$ cm$^{-3}$) & 5.62\\ [0.25ex]
Diffuse Hydrogen density ($\times$ 10$^{8}$ cm$^{-3}$) & 4.46\\ [0.25ex]
Covering factor (clump) & 0.40, 0.40$^{a}$\\ [0.25ex]
Covering factor (diffuse) & 0.20\\ [0.25ex]
$\alpha$ &-3.00\\ [0.25ex]
Inner radius ($\times$ 10$^{14}$ cm) & 3.63\\ [0.25ex]
Outer radius ($\times$ 10$^{15}$ cm) & 1.00\\ [0.25ex]
Filling factor & 0.1\\ [0.25ex]
Fe/Fe$_{\odot}$ & 1.41 (12)$^{b}$\\ [0.25ex]
N/N$_{\odot}$ & 5.76 (5)\\ [0.25ex]
Ca/Ca$_{\odot}$ & 1.13 (2)\\ [0.25ex]
C/C$_{\odot}$ & 1.00 (1)\\ [0.25ex]
He/He$_{\odot}$ & 2.00 (4)\\ [0.25ex]
Ejected mass ($\times$ 10$^{-5}$ M$_{\odot}$) & 2.74\\ [0.25ex]
Number of observed lines (n) & 27\\ [0.25ex]
Number of free parameters (n$_{p}$) & 12\\ [0.25ex]
Degrees of freedom ($\nu$) & 15\\ [0.25ex]
Total $\chi^{2}$ & 23.08 \\ [0.25ex]
$\chi^{2}_{red}$ & 1.54\\ [0.25ex]
\hline
\end{tabular}}
\end{center}
$^{a}$The covering factor used for the third region to fit N II lines.
$^{b}$The number in the parenthesis is the number of lines used to determine abundance estimate.
\end{table}

\subsection{Nebular phase}
Three epochs i.e., day 224, 252 and 336 were modelled to understand the evolution during the nebular phase. Two regions were used to model the observed spectrum. Most of the lines were fitted by the clump component, except a few forbidden lines like [N II] (5755 \AA), [O III] (4959 \& 5007 \AA) and [Ne III] (3869 \& 3968 \AA). In order to fit all the lines, a diffuse region (low density) was used, covering 20\% of the volume. The central ionizing source was set to be at an effective temperature $\sim$ 10$^{5}$ K and luminosity $\sim$ 10$^{37}$ erg s$^{-1}$. The clump hydrogen density is in the range of (5.6--7.9) $\times$ 10$^{7}$ cm$^{-3}$ and diffuse hydrogen density, (1.8--3.2) $\times$ 10$^{7}$ cm$^{-3}$. The relative fluxes of the observed lines, best-fit model predicted lines and corresponding $\chi^{2}$ values are given in \Cref{cval_nq}. The values of best-fit parameters obtained from model are given in \Cref{param_n}. The estimated abundance values show that helium, nitrogen, oxygen and neon are over abundant compared to solar, while iron and carbon have the solar abundance values. On day 252, nitrogen and helium abundance values are more than that of day 224 and 336. Oxygen abundance values continue to be above the solar values in all the epochs. The best-fit modelled spectra for all the epochs are shown in \Cref{ccc} together with the corresponding observed spectra. 
\begin{table*}
\caption{Observed and best-fit CLOUDY model line flux values$^{a}$ for epochs of nebular and late post-outburst phase}
\label{cval_nq}
\begin{center}
\resizebox{1\textwidth}{!}
{\begin{tabular}{l c c c c c c c c c c c c c} 
 \hline
  \textbf{Line ID} & \boldmath{$\lambda$ (\AA)} & \textbf{Observed} & \textbf{modelled} & \boldmath{$\chi^{2}$} & \textbf{Observed} & \textbf{modelled} & \boldmath{$\chi^{2}$} & \textbf{Observed} & \textbf{modelled} & \boldmath{$\chi^{2}$} & \textbf{Observed} & \textbf{modelled} & \boldmath{$\chi^{2}$}\\ 
  & & \multicolumn{3}{c}{Day 224} & \multicolumn{3}{c}{Day 252} & \multicolumn{3}{c}{Day 336} & \multicolumn{3}{c}{Day 1064}\\ [0.5ex] 
 \hline\hline
 [Ne III] & 3869 & 5.51E$-$01 & 4.73E$-$01 & 1.00E$-$01$^{b}$ & 8.96E$-$01 & 6.33E$-$01& 7.24E$-$01 & ... & ... & ... & ... & ... & ...\\ [0.25ex]
[Ne III] & 3968 & 7.65E$-$02 & 1.08E$-$01 & 3.67E$-$02 & 3.09E$-$01 & 3.83E$-$01 & 2.18E$-$01 & ... & ... & ... & ... & ... & ...\\ [0.25ex]
H I & 4102 & 3.58E$-$01 & 3.51E$-$01 & 2.17E$-$03 & 1.16E$+$00 & 9.19E$-$01 & 7.31E$-$01 & 8.12E$-$01 & 6.97E$-$01 & 4.02E$-$01 & 4.94E$-$01 & 4.39E$-$01 & 2.61E$-$02\\ [0.25ex]
H I, [C III], [O III] & 4364 & 3.54E$+$00 & 3.52E$+$00 & 6.45E$-$02 & 6.23E$+$00 & 5.90E$+$00 & 1.22E$+$00 & 3.47E$+$00 & 3.19E$+$00 & 1.46E$+$00 & 1.19E$+$00 & 8.95E$-$01 & 1.22E$+$00\\ [0.25ex]
N III & 4517 & 2.23E$-$01 & 6.14E$-$02 & 2.58E$-$01 & 3.02E$-$01 & 3.12E$-$02 & 4.59E$-$01 & 3.47E$-$01 & 1.68E$-$01 & 4.98E$-$01 & ... & ... & ...\\ [0.25ex]
N III + C III & 4640 & 1.07E$+$00 & 7.84E$-$02 & 1.04E$+$00 & 1.50E$+$00 & 1.32E$+$00 & 5.79E$-$01 & 5.57E$-$01 & 4.55E$-$01 & 1.83E$-$01 & 5.64E$-$01 & 7.58E$-$01 & 8.16E$-$01\\ [0.25ex]
He II & 4686 & 5.42E$-$01 & 2.78E$-$01 & 1.66E$+$00 & 8.59E$-$01 & 6.32E$-$01 & 7.04E$-$01 & 4.45E$-$01 & 5.76E$-$01 & 1.15E$+$00 & 1.79E$+$00 & 2.06E$+$00 & 0.80E$-$01\\ [0.25ex]
H I & 4861 & 1.00E$+$00 & 1.00E$+$00 & 0.00E$+$00 & 1.00E$+$00 & 1.00E$+$00 & 0.00E$+$00 & 1.00E$+$00 & 1.00E$+$00 & 0.00E$+$00 & 1.00E$+$00 & 1.00E$+$00 & 0.00E$+$00\\ [0.25ex]
[O III] & 4959 & 3.69E$+$00 & 3.96E$+$00 & 7.95E$-$01 & 3.17E$+$00 & 3.33E$+$00 & 3.14E$-$01 & 6.87E$+$00 & 7.31E$+$00 & 2.17E$+$00 & 2.84E$+$00 & 2.53E$+$00 & 1.41E$+$00\\ [0.25ex]
[O III] & 5007 & 9.27E$+$00 & 9.36E$+$00 & 1.01E$-$01 & 1.39E$+$01 & 1.36E$+$01 & 
1.09E$+$00 & 2.42E$+$01 & 2.41E$+$01 & 9.41E$-$02 & 7.08E$+$00 & 6.74E$+$00 & 9.65E$-$01\\ [0.25ex]
[Fe VII] & 5158 & 3.75E$-$01 & 6.33E$-$01 & 4.48E$+$00 & 2.68E$-$01 & 4.63E$-$03 & 3.46E$+$01 & ... & ... & ... & ... & ... & ...\\ [0.25ex]
[Fe VII] & 5276 & 2.66E$-$01 & 1.97E$-$02 & 6.56E$-$01 & 2.57E$-$01 & 7.53E$-$03 & 9.95E$-$01 & ... & ... & ... & ... & ... & ...\\ [0.25ex]
He II & 5411 & 1.17E$-$01 & 2.38E$-$02 & 1.19E$-$01 & 3.64E$-$01 & 1.79E$-$02 & 1.83E$-$03 & ... & ... & ... & 2.75E$-$01 & 2.52E$-$01 & 5.78E$-$03\\ [0.25ex]
N II & 5679 & ... & ... & ... & 3.53E$-$02 & 1.12E$-$02 & 1.05E$-$01 & ... & ... & ... & ... & ... & ...\\ [0.25ex]
[N II] & 5755 & 8.28E$-$01 & 9.87E$-$01 & 1.01E$+$00 & 7.47E$-$01 & 4.71E$-$01 & 1.00E$+$00 & 6.78E$-$01 & 4.07E$-$01 & 1.46E$+$00 & ... & ... & ...\\ [0.25ex]
He I & 5876 & 2.20E$-$01 & 1.05E$-$01 & 2.37E$-$01 & 5.32E$-$02 & 2.14E$-$01 & 8.06E$-$01 & 1.19E$-$01 & 1.85E$-$01 & 5.69E$-$01 & 4.53E$-$01 & 1.71E$-$01 & 8.74E$-$01\\ [0.25ex]
[Fe VII] & 6087 & 9.91E$-$02 & 2.16E$-$03 & 2.50E$-$01 & 3.23E$-$01 & 1.84E$-$02 & 1.51E$+$00 & 5.51E$-$01 & 4.34E$-$01 & 2.34E$-$01 & ... &  ... & ...\\ [0.25ex]
H I + [N II] & 6563 & 1.34E$+$00 & 1.06E$+$00 & 1.18E$+$00 & 1.62E$+$00 & 1.93E$+$00 & 1.18E$+$00 & 2.41E$+$00 & 2.66E$+$00 &1.06E$+$00 & 3.50E$+$00 & 3.14E$+$00 & 1.36E$+$00\\ [0.25ex]
He I & 6678 & 1.45E$-$02 & 4.52E$-$02 & 1.47E$-$02 & 1.70E$-$02 & 5.80E$-$02 & 1.29E$-$02 & ... & ... & ... & 2.66E$-$01 &2.89E$-$01 & 2.77E$-$02\\ [0.25ex]
He I & 7065 & ... & ... & ... & ... & ... & ... & ... & ... & ... & 3.88E$-$01 & 9.05E$-$02 & 1.17E$+$00\\ [0.25ex]
[O II] & 7330 & ... & ... & ... & 9.51E$-$02 & 1.18E$-$01 & 6.02E$-$02 & 2.19E$-$01 & 4.32E$-$02 & 4.59E$-$01 & ... & ... & ...\\ [0.25ex]
\hline
\end{tabular}}
\end{center}
$^{a}$ Relative to H$\beta$ \\
$^{b}$ Calculated using equation \Cref{eqn2}
\end{table*}

\begin{table}
\caption{Best-fit CLOUDY model parameters during the nebular phase}
\label{param_n}
\begin{center}
\resizebox{\columnwidth}{!}{%
\begin{tabular}{l c c c} 
 \hline
  \textbf{Parameter} & \textbf{Day 224} & \textbf{Day 252} & \textbf{Day 336} \\ [0.5ex] 
 \hline\hline 
 T$_{BB}$ ($\times$ 10$^{5}$ K) & 1.50 & 2.5 & 3\\ [0.25ex]
Luminosity ($\times$ 10$^{37}$ erg/s) & 3.16 & 3.98 & 5.01 \\ [0.25ex]
Clump Hydrogen density ($\times$ 10$^{6}$ cm$^{-3}$) & 79.43 & 63.09 & 56.23 \\ [0.25ex]
Diffuse Hydrogen density ($\times$ 10$^{6}$ cm$^{-3}$) & 31.62 & 25.12 & 17.78 \\ [0.25ex]
Covering factor (clump) & 0.80 & 0.80 & 0.80 \\ [0.25ex]
Covering factor (diffuse) & 0.20 & 0.20 & 0.20 \\ [0.25ex]
$\alpha$ & -3.00 & -3.00 & -3.00 \\ [0.25ex]
Inner radius ($\times$ 10$^{15}$ cm) & 2.34 & 2.88 & 3.09 \\ [0.25ex]
Outer radius ($\times$ 10$^{16}$ cm) & 0.38 & 0.85 & 1.26 \\ [0.25ex]
Filling factor  & 0.1 & 0.1 & 0.1 \\ [0.25ex]
He/He$_{\odot}$ & 1.43 (4)$^{a}$ & 1.65 (4) & 1.16 (2) \\ [0.25ex]
N/N$_{\odot}$ & 3.02 (4) & 3.63 (5) & 2.96 (4) \\ [0.25ex]
O/O$_{\odot}$ & 3.01 (3) & 2.65 (3) & 2.63 (4) \\ [0.25ex]
Ne/Ne$_{\odot}$ & 3.25 (2) & 1.36 (2) & ... \\ [0.25ex]
C/C$_{\odot}$ & 1.00 (1) & 1.02 (1) & 1.09 (1) \\ [0.25ex]
Fe/Fe$_{\odot}$ & 1.09 (3) & 1.00 (3) & 1.00 (1) \\ [0.25ex]
Ejected mass ($\times$ 10$^{-6}$ M$_{\odot}$) & 0.46 & 1.52 & 2.13 \\ [0.25ex]
Number of observed lines (n) & 18 & 20 & 13 \\ [0.25ex]
Number of free parameters (n$_{p}$) & 10 & 10 & 7 \\ [0.25ex]
Degrees of freedom ($\nu$) & 8 & 10 & 6 \\ [0.25ex]
Total $\chi^{2}$ & 12.74 & 15.18 & 9.22 \\ [0.25ex]
$\chi^{2}_{red}$ & 1.59 & 1.52 & 1.54 \\ [0.25ex]
\hline
\end{tabular}}
\end{center}
$^{a}$The number in the parenthesis is the number of lines used to determine abundance estimate.
\end{table}

\begin{figure*}
\centering
   \includegraphics[width=11cm]{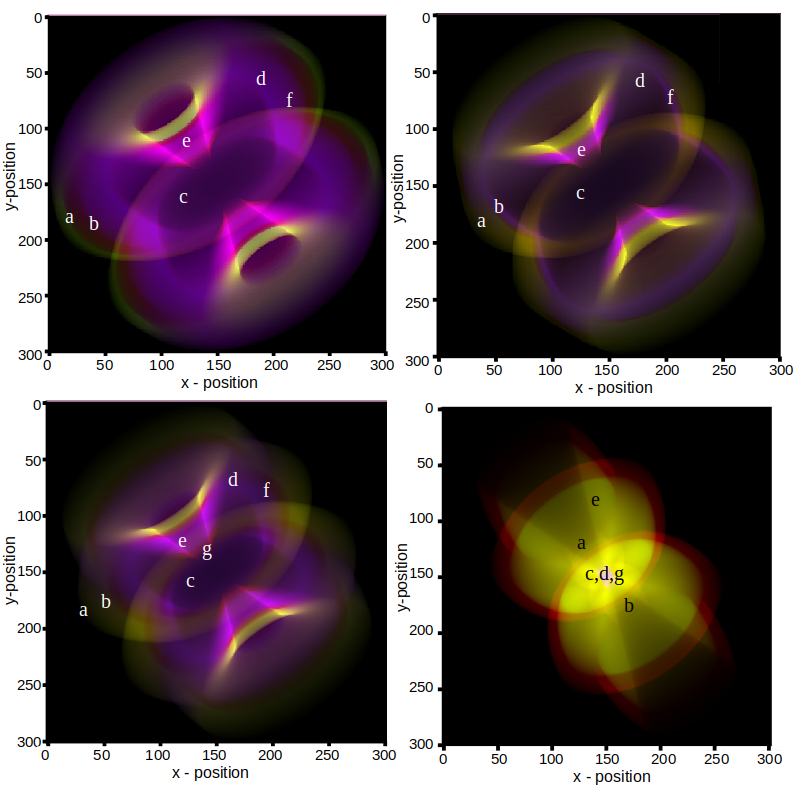}
     \caption{The evolution of the geometry of ejecta on day 224 (top left), 252 (top right), 336 (bottom left) and 1064 (bottom right). Notice the formation of multiple equatorial rings and also the spatial distribution of the ionized lines as the system evolves. Here, a: [O III] $\lambda$5007, b: [O III] $\lambda$4959, c: H$\alpha$, d: H$\beta$, e: N III $\lambda$4638, f: [N II] $\lambda$5755, g: He II $\lambda$4686. Only dominant lines are marked. Here, 1 unit of x and y position correspond to x$_{224}$ = 6.3 $\times$ 10$^{12}$ cm, x$_{252}$ = 1.41 $\times$ 10$^{13}$ cm, x$_{336}$ = 2.09 $\times$ 10$^{13}$ cm and x$_{1064}$ = 1.66 $\times$ 10$^{12}$ cm.}
     \label{3dmodel}
\end{figure*}

\begin{figure*}
\centering
   \includegraphics[width=11cm]{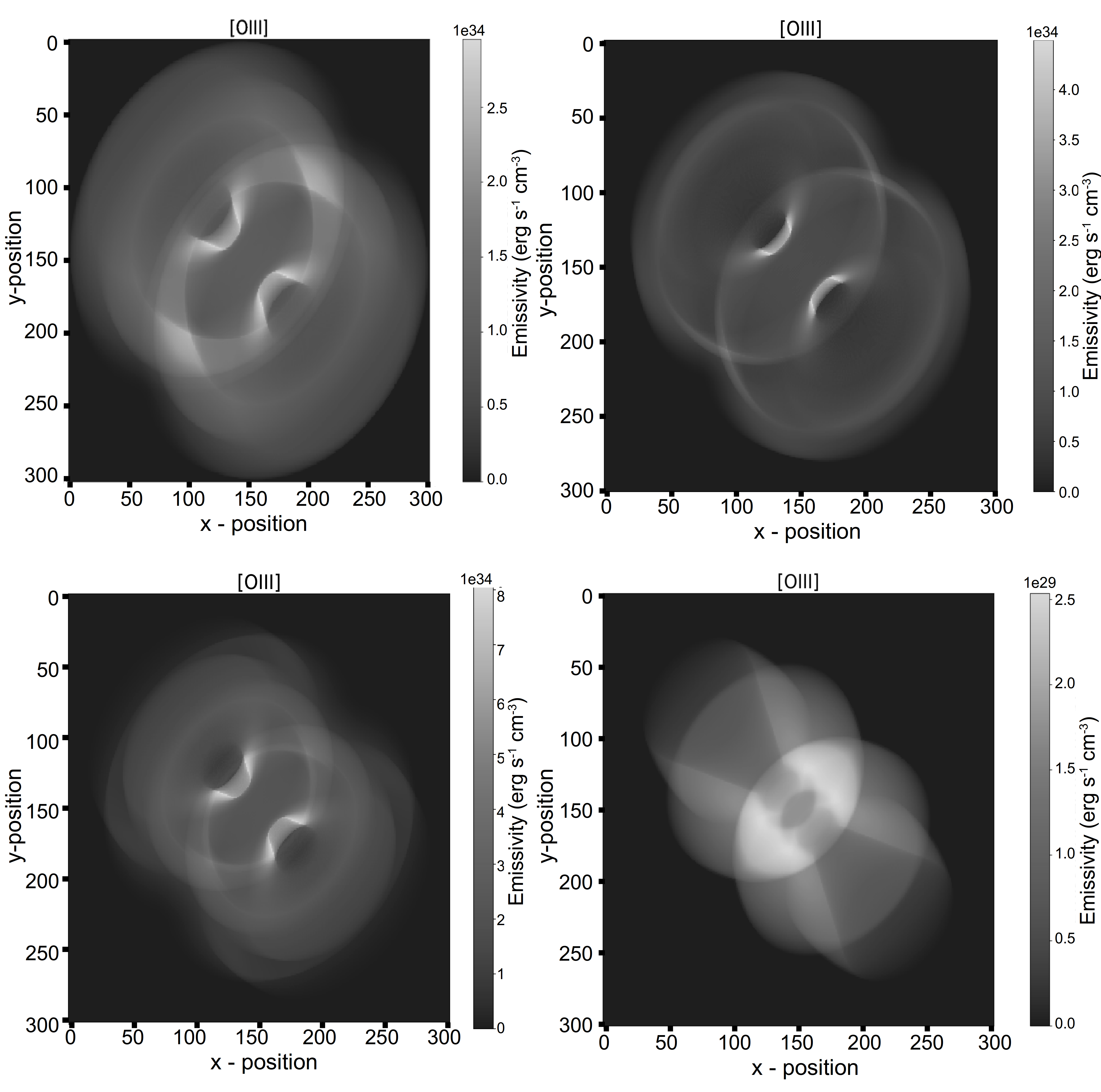}
     \caption{The evolution of [O III] (4959 \& 5007 \AA\ lines) on day 224 (top left), 252 (top right), 336 (bottom left) and 1064 (bottom right). Notice the increase in intensity from day 224 to day 336 and then drop in intensity on day 1064. Here, 1 unit of x and y position correspond to x$_{224}$ = 6.3 $\times$ 10$^{12}$ cm, x$_{252}$ = 1.41 $\times$ 10$^{13}$ cm, x$_{336}$ = 2.09 $\times$ 10$^{13}$ cm and x$_{1064}$ = 3.31 $\times$ 10$^{11}$ cm.}
     \label{o3model}
\end{figure*}

The geometry of the ionized structure during this phase was found to be a bipolar conical one with equatorial rings (\Cref{3dmodel}). The [O III] emission is dominant in the outermost regions of the cones and equatorial rings. The emissivity increases with time while the structure remains the same (\Cref{o3model}). The nitrogen lines are present in the innermost regions of equatorial rings and cones. The He II (4686 \AA) is present in the innermost regions of cones, while hydrogen lines are present in the innermost regions of the equatorial rings. No significant change is seen in the emissivity of these lines.


\subsection{Late post-outburst phase}
During this phase, spectrum obtained on day 1064 was modelled. The observed spectrum is dominated by hydrogen, helium and [O III] lines as discussed in \Cref{2.2}. A two component model was considered in order to generate the model spectrum. One component with a luminosity of 10$^{35}$ erg s$^{-1}$ as an ionizing source with disc of density, 2 $\times$ 10$^{9}$ cm$^{-3}$ at T = 7.5 $\times$ 10$^{4}$ K and cylindrical semi height, 10$^{4}$ cm. The helium and hydrogen lines were dominantly fit by this component. The other component with similar ionizing source luminosity and lesser density region, 10$^{6}$ cm$^{-3}$ at T = 9.5 $\times$ 10$^{4}$ K. The [O III] and nitrogen lines were fit using this component. The radius, luminosity and density were calculated by adopting the values of mass of the WD, accretion rate, period and binary separation from \citet{Pat17} and \citet{Sel08}. The modelled spectrum was obtained by adding the values generated by the above components (\Cref{ccc}). Best-fit parameters were obtained for these values (\Cref{param_q}). The estimated abundance values of helium, nitrogen and oxygen were found to be more than that of the solar abundance values. 

\begin{table}
\caption{Best-fit CLOUDY model parameters for the late post-outburst phase spectrum}
\label{param_q}
\begin{center}
\resizebox{\columnwidth}{!}{%
\begin{tabular}{l c c c c} 
 \hline
  \textbf{Parameter} &  \textbf{Component 1/Component 2}\\[0.5ex]
 \hline\hline
T$_{BB}$ ($\times$ 10$^{4}$ K) & 7.5/9.5\\[0.25ex]
Luminosity (log(L in erg/s)) & 35/35\\ [0.25ex]
Hydrogen density & 9.3/6\\ [0.25ex]
$\alpha$ & -3.00\\ [0.25ex]
Inner radius (log(r in cm)) & 9/12\\ [0.25ex]
Outer radius (log(r in cm)) & 12/15\\ [0.25ex]
Covering factor & 0.65/0.35\\ [0.25ex]
Filling factor  & 0.1\\ [0.25ex]
He/He$_{\odot}$ & 5.35 (5)$^{a}$\\ [0.25ex]
N/N$_{\odot}$ & 2.78 (2)\\ [0.25ex]
O/O$_{\odot}$ & 2.21 (2)\\ [0.25ex]
Shell mass ($\times$ 10$^{-6}$ M$_{\odot}$) & 3.64 \\ [0.25ex]
Number of observed lines (n) & 12\\ [0.25ex]
Number of free parameters (n$_{p}$) & 7\\ [0.25ex]
Degrees of freedom ($\nu$) & 5\\ [0.25ex]
Total $\chi^{2}$ & 8.69\\ [0.25ex]
$\chi^{2}_{red}$ & 1.74\\ [0.25ex]
\hline
\end{tabular}}
\end{center}
$^{a}$The number in the parenthesis is the number of lines used to determine abundance estimate.
\end{table}
The ionized structures of different lines obtained in this phase have distinct spatial location unlike previous epochs. The overall geometry of the ionized ejecta is dominantly divided into nitrogen zone, oxygen zone and helium-hydrogen zone (from outer to inner regions) as shown in \Cref{3dmodel}. The conical structures appear have evolved compared to the previous epochs. The outermost regions and the cones are of nitrogen lines and the next inner regions are of [O III] lines indicating that they are coming from shell ejected by the system. The helium lines are coming from the region of a radius of about 7.4 $\times$ 10$^{9}$ cm. If the accretion rate suggested by \cite{God18} is considered, then this radius value would correspond to that of the outermost regions of accretion disc. This suggests that helium lines are predominantly coming from the accretion disc. 

\subsection{Results}
\begin{figure}
\centering
   \resizebox{\hsize}{!}{\includegraphics{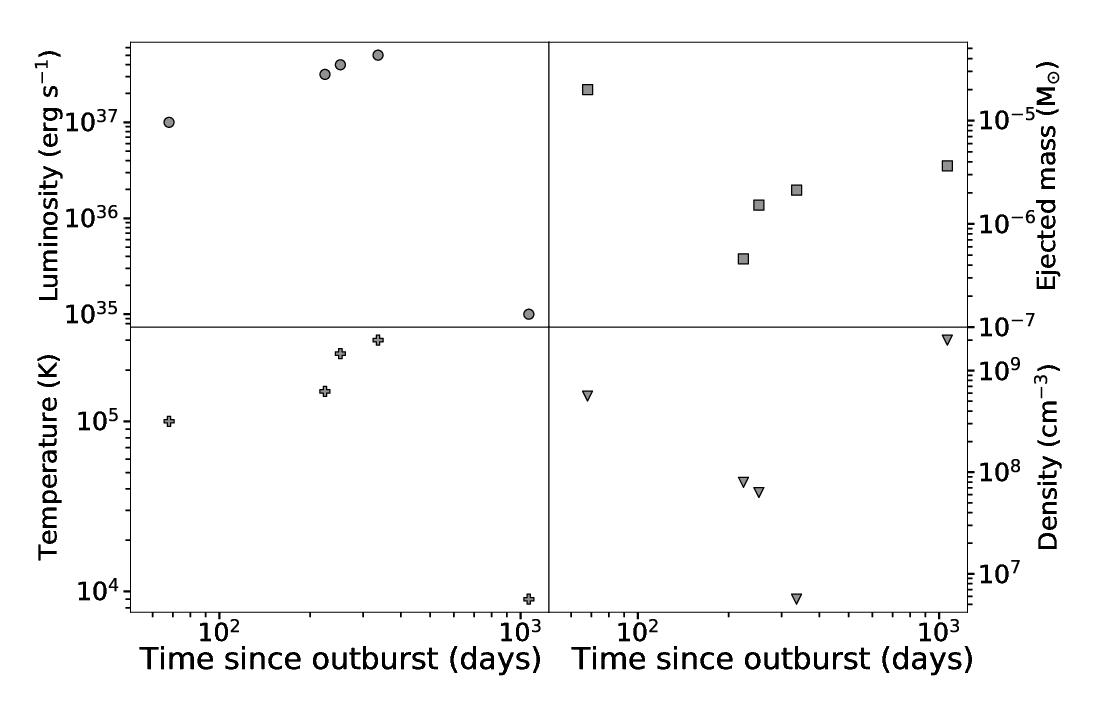}}
     \caption{The variation of best-fit parameters obtained from CLOUDY such as luminosity, effective temperature, density and ejected shell mass are shown.}
     \label{param}
\end{figure}

Best-fit model obtained for day 68 was used to fit only the emission components. The P-Cygni profiles and O I (7774 \AA) line were not modelled due to the limitations of the code. N II (5679 \AA) line has high $\chi^{2}$ value, it has a broad emission component and narrow absorption component. He II (4686 \AA) line has higher $\chi^{2}$ values and hence contribute more to total $\chi^{2}$. In the late post-outburst phase, [O III] and He I (7065 \AA) lines have higher $\chi^{2}$ values. Though the reduced $\chi^{2}$ values suggest that the generated spectrum match the observed spectrum well, above points might hinder the calculated abundance values. The higher $\chi^{2}$ values are as a consequence of blending of the lines. The [Fe VII] (6087 \AA) line has higher $\chi^{2}$ value on few days. It is a coronal line possibly excited by shock interaction, hence resulting in a higher $\chi^{2}$ value. The variation of best-fit parameters during the system's evolution is given in \Cref{param}.

The estimate of the ejected mass, averaged over all epochs, is $7.03 \times\, 10^{-6}\, M_{\odot}$. This value is similar to that obtained by \citet{Sho13}. Best-fit model parameters suggest the presence of a hot WD source with a roughly constant luminosity of $10^{37}$ erg s$^{-1}$. Helium and nitrogen abundance values are above solar values in all the phases, whereas neon is more only during nebular phase. Oxygen abundance is also found to be more than solar, while the iron and calcium abundances are nearly solar. Though the distance to T Pyx used for analysis is 4.8 kpc, the results were also verified for the distance given by Gaia i.e., $\sim$3 kpc. The 1D CLOUDY results do not vary as relative intensity is considered for the calculations. 

\begin{figure}
\centering
   \resizebox{\hsize}{!}{\includegraphics{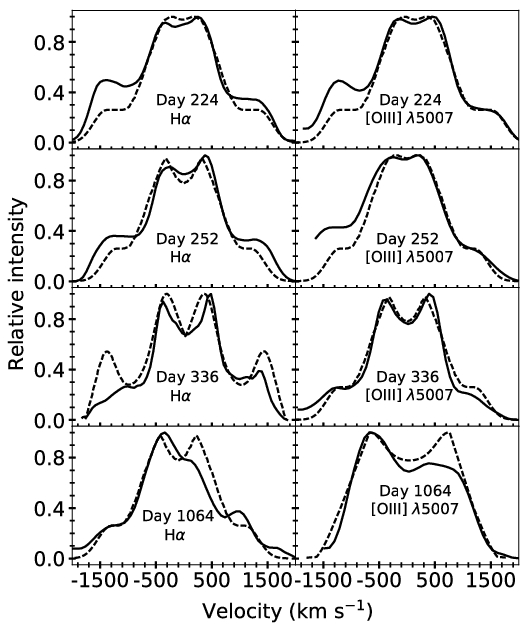}}
     \caption{Best-fit pyCloudy model velocity profile (dash line) at i $\sim$ 14.75$^\circ$ plotted over the observed profile of H$\alpha$ and [O III] $\lambda$5007 (continuous line) of T Pyx obtained on day 224, 252, 336 and 1064.}
     \label{all_pyc}
\end{figure}

The modelled structure of the ionized ejecta shows that the evolution of the ejecta is consistent with the line profiles observed. The variation in the spatial distribution of the elements as the physical conditions change from one epoch to another is consistent with the observed spectraThe best-fit modelled velocity profile (dash line) with the corresponding observed optical velocity profile (continuous line) of H$\alpha$ and [O III] $\lambda$5007 are as shown in \Cref{all_pyc}. The best-fit modelled profiles are symmetric here as the initial geometry was assumed to be spherically symmetric (refer introduction of \Cref{sec4}). The inner and outer cone angles found using this method are consistent with that estimated by \cite{Sho13}. The inner conical angles lie in the range of 21.08$^\circ$ to 77.86$^\circ$ while the outer conical angles varied from 49.89$^\circ$ to 87.75$^\circ$. The value of inclination angle of ejecta axis to line-of-sight was found to be $14.75^\circ\, \pm\, 0.65^\circ$. This value is consistent with that of the values reported by \citet{Pat98,Che11,Sho13}.

\section{Discussion} 
The optical spectral evolution of T Pyx from its 2011 outburst discovery to the subsequent quiescent phase has been discussed in this paper. This outburst is similar to that of the previous ones, but many details relevant to the system like variation in physical conditions at different epochs i.e., presence of distinct ionization lines, evolution of line profiles were possible to study because of good coverage.

Spectral lines observed have P-Cygni profiles with sharp and deep narrow absorption components and narrow emission components till the early decline phase. These slowly evolve into broader double-peaked emission lines with wing-like structures during the nebular phase. Finally, these wing-like structures (left and right) disappear resulting in sharp and narrow emission lines in the late post-outburst phase. There is decrease in ejecta velocity from $\sim$ 2500 km s$^{-1}$ to $\sim$ 1000 km s$^{-1}$ during the initial pre-maximum phase from t = 1.28 to 12.32 and then velocity increases up to $\sim$ 2000 km s$^{-1}$ on its way to optical maximum and decline phase. Weak N II lines from day 14.31 are seen which become stronger towards the decline phase. Transition of the system from He/N to Fe II type occurs when it is rising to its optical maximum. [O III] 4959 and 5007 \AA\ lines appear to be blue-shifted in the late post-outburst phase while other lines are not shifted. This might be due to motion of the doubly ionized oxygen shell towards us. 

Nearly solar abundances are found for iron and calcium seen during the optically thick Fe II phase, while elements like helium, nitrogen, oxygen and neon are more than solar during both the Fe II and nebular (He/N) phases. The ejected mass during the decline phase at t = 68 days is estimated to be $2.74\, \times\, 10^{-5}\, M_{\odot}$. During the nebular phase, it is of the order of $10^{-6}\, M_{\odot}$. Based on the observed period change before and after the 2011 eruption, \citet{Pat17} estimated the ejected mass to be $\ge 3\, \times\, 10^{-5}\, M_{\odot}$, while \citet{Nel14} estimated the ejected mass as $(1-30)\, \times\, 10^{-5}\, M_{\odot}$ using the high peak flux densities in the radio emission. Based on the late turn-on time for the SSS phase, \citet{cho14} suggested a large ejecta mass of $\gtrapprox\, 10^{-5}\, M_{\odot}$. Using high resolution optical spectroscopic observations obtained on day 180 and 360, \citet{Sho13} estimated the ejected mass to be $M_{\rm{ej}}=2\, \times\, 10^{-6}\, M_\odot$, similar to the values reported here for the nebular phase. The ejected mass estimates for other recurrent novae are $\sim 10^{-7} - 10^{-6}\, M_{\odot}$, while the estimates for classical novae are, in general, higher at $\sim 10^{-5} - 10^{-4}\, M_{\odot}$. The ejected mass obtained for the T Pyx system in the early phase (day 68) is similar to classical novae, while for the nebular phase the ejected mass values are similar to recurrent novae. Using accretion disk models, \cite{God18} estimate a mass accretion rate of $10^{-6}\, M_{\odot} \rm{yr}^{-1}$ for a distance of 4.8 kpc, or a rate of  $10^{-7}\, M_{\odot} \rm{yr}^{-1}$ for the Gaia distance estimate of $\sim$ 3.3 kpc, for a white dwarf mass of $1.35\, M_{\odot}$.  This indicates the mass accumulated between the 1967 and 2011 outbursts is lower than the ejected mass, for an ejecta mass of $\ga 10^{-5}\, M_{\odot}$.

The geometry of ionized structure of the ejecta is a bipolar conical one with equatorial rings. It is noticed that [O III] and nitrogen lines are mainly coming from the outer regions, and hydrogen lines from the inner regions. As the shell evolves, expansion of equatorial rings can clearly be seen in the nebular phase. In the late post-outburst phase, a complex structure of [O III] and nitrogen are formed with an inner helium and hydrogen structures, most likely from the accretion disc. From the study of \citet{Che11}, it is known that the system is oriented nearly face-on, and it is bipolar. Hence, the outer regions (especially [O III]) are moving away from the central system towards us (nearly face-on) resulting in the blue-shifted [O III] lines from day 1721.

Polarimetric studies of novae have revealed the presence of linear polarization in systems like U Sco \citep{gca13}, V339 Del \citep{Sha17}, RS Oph \citep{Cro90} and also during the 1967 outburst of T Pyx \citep{Egg67}. The intrinsic linear polarization during the SSS phase in U Sco was explained as due to scattering from discs and jets. In the case of V339 Del, variability in the polarization parameters was observed, and interpreted as being due to a non-spherical diffuse shell, with a geometry that was more likely bipolar than disc-shaped. In RS Oph (1985 outburst), the intrinsic linear polarization was suggested as due to electron scattering in the aspherically expanding nova ejecta. Linear polarized emission has been observed from the T Pyx system in BVRI bands during the 1967 as well as the 2011 outbursts. The observed polarization in T Pyx is variable, with a very similar trend during both outbursts. Although not corrected for interstellar polarization the similar trend indicates the polarization to be intrinsic, which can suggested as being due to a) asymmetry of the ejecta at the time of outburst, and/or b) presence of silicate grains, as suggested by \cite{svatovs1983intrinsic} for the 1967 eruption. IR observations of T Pyx during the 2011 outburst suggest the presence of a pre-outburst dust in the system with a mass of $\sim10^{-5}\, M_{\odot}$ \citep{Eva12}. Asymmetry in the system at the time of outburst can be due to interaction of the initial ejecta with circumstellar material, also resulting in the decrease in velocity as observed in the spectral data till day 12.32. Coincidentally, there was a marginal detection of hard X-rays during days 14-20, which has been attributed by \cite{cho14} to an interaction of the nova ejecta with a pre-existing circumbinary material. Such an interaction has also been shown to result in equatorial rings / regions \citep{Soker15, Bod07, Liv90, Llo97} similar to that seen in the models presented here.

\section{Summary}
\begin{enumerate}
\item Spectral evolution of the 2011 outburst of the recurrent nova T Pyx is reported here. The evolution is similar to that of the previous outbursts.
\item The emission lines during the early pre-maximum phase have P-Cygni profiles with deep and narrow absorption components and sharp emission peaks, with the ejecta velocity decreasing from 2500 km s$^{-1}$ to 1000 km s$^{-1}$ till $\sim$day 12.
\item The rise to optical maximum and early decline phase have P-Cygni profiles with round peaked emission components and slowly fading absorption components with the ejecta velocity increasing from 1000 km s$^{-1}$ to 2000 km s$^{-1}$ till $\sim$day 68.
\item The emission lines in the nebular phase are broad and double peaked with wing-like structures on both the sides. The emission lines in late post-outburst phase are narrow, and [O III] lines show a blueshift by $\sim$ 780 km s$^{-1}$ beyond day 1721.
\item The photo-ionization code, CLOUDY was used to model the spectra of five epochs, and the modelled spectra match the observational spectra fairly well. From these results, elemental abundances, temperature, luminosity of the WD and density of the ejecta were estimated at all the epochs. Helium, nitrogen, oxygen and neon were found to be over abundant compared to solar values.
\item The ejected mass during the Fe II phase of the system is estimated to be $2.74\, \times\, 10^{-5}\, M_{\odot}$ while during the nebular phase it is $\sim 2\, \times\, 10^{-6}\, M_{\odot}$.
\item The pseudo 3D code, pyCLOUDY was used to model the ionized structure of ejecta for four epochs. The geometry was found to be bipolar conical with equatorial rings, with an inclination angle of $14.75^\circ\, \pm\, 0.65^\circ$. At late post-outburst phase, it appears that the [O III] lines come from an expanding ejecta while the hydrogen and helium lines are from the accretion disc. 
\item A variable, intrinsic linear polarization was observed, which could be either due to asymmetry in the initial ejecta and/or presence of silicate grains.
\end{enumerate}
\begin{acknowledgements}
      We thank the referee for a critical reading of the manuscript. We thank all the observers of the 2 m HCT at the IAO, 2 m IGO telescope at the IGO and also the 2.3 m VBT at the VBO for accommodating some time for ToO observations. We thank the respective TACs for the allocation of time and support during ToO and regular observations. The IAO and VBO are operated by the Indian Institute of Astrophysics, Bangalore and the IGO by Inter-University Centre for Astronomy and Astrophysics, Pune. We thank Mr. Avinash Singh for the aperture photometry code\footnote{https://github.com/sPaMFouR/RedPipe} which was used for polarization data reduction. 
\end{acknowledgements}



\begin{thebibliography}{48}
	\expandafter\ifx\csname natexlab\endcsname\relax\def\natexlab#1{#1}\fi
	
	\bibitem[{{Anupama}(2008)}]{gca08}
	{Anupama}, G.~C. 2008, in Astronomical Society of the Pacific Conference
	Series, Vol. 401, RS Ophiuchi (2006) and the Recurrent Nova Phenomenon, ed.
	A.~{Evans}, M.~F. {Bode}, T.~J. {O'Brien}, \& M.~J. {Darnley}, 31
	
	\bibitem[{{Anupama} {et~al.}(2013){Anupama}, {Kamath}, {Ramaprakash},
		{Kantharia}, {Hegde}, {Mohan}, {Kulkarni}, {Bode}, {Eyres}, {Evans}, \&
		{O'Brien}}]{gca13}
	{Anupama}, G.~C., {Kamath}, U.~S., {Ramaprakash}, A.~N., {et~al.} 2013, \aap,
	559, A121
	
	\bibitem[{{Bode} {et~al.}(2007){Bode}, {Harman}, {O'Brien}, {Bond},
		{Starrfield}, {Darnley}, {Evans}, \& {Eyres}}]{Bod07}
	{Bode}, M.~F., {Harman}, D.~J., {O'Brien}, T.~J., {et~al.} 2007, \apjl, 665,
	L63
	
	\bibitem[{{Chesneau} {et~al.}(2011){Chesneau}, {Meilland}, {Banerjee}, {Le
			Bouquin}, {McAlister}, {Millour}, {Ridgway}, {Spang}, {ten Brummelaar},
		{Wittkowski}, {Ashok}, {Benisty}, {Berger}, {Boyajian}, {Farrington},
		{Goldfinger}, {Merand}, {Nardetto}, {Petrov}, {Rivinius}, {Schaefer},
		{Touhami}, \& {Zins}}]{Che11}
	{Chesneau}, O., {Meilland}, A., {Banerjee}, D.~P.~K., {et~al.} 2011, \aap, 534,
	L11
	
	\bibitem[{{Chomiuk} {et~al.}(2014){Chomiuk}, {Nelson}, {Mukai}, {Sokoloski},
		{Rupen}, {Page}, {Osborne}, {Kuulkers}, {Mioduszewski}, {Roy}, {Weston}, \&
		{Krauss}}]{cho14}
	{Chomiuk}, L., {Nelson}, T., {Mukai}, K., {et~al.} 2014, \apj, 788, 130
	
	\bibitem[{{Clarke} {et~al.}(1998){Clarke}, {Smith}, \& {Yudin}}]{Cla98}
	{Clarke}, D., {Smith}, R.~A., \& {Yudin}, R.~V. 1998, \aap, 336, 604
	
	\bibitem[{{Cropper}(1990)}]{Cro90}
	{Cropper}, M. 1990, \mnras, 243, 144
	
	\bibitem[{{Das} \& {Mondal}(2015)}]{Das15}
	{Das}, R. \& {Mondal}, A. 2015, \na, 39, 19
	
	\bibitem[{{De Gennaro Aquino} {et~al.}(2014){De Gennaro Aquino}, {Shore},
		{Schwarz}, {Mason}, {Starrfield}, \& {Sion}}]{deg14}
	{De Gennaro Aquino}, I., {Shore}, S.~N., {Schwarz}, G.~J., {et~al.} 2014, \aap,
	562, A28
	
	\bibitem[{{Duerbeck}(1987)}]{Due87}
	{Duerbeck}, H.~W. 1987, \ssr, 45, 1
	
	\bibitem[{{Duerbeck} \& {Seitter}(1979)}]{Due79}
	{Duerbeck}, H.~W. \& {Seitter}, W.~C. 1979, The Messenger, 17, 1
	
	\bibitem[{{Eggen} {et~al.}(1967){Eggen}, {Mathewson}, \& {Serkowski}}]{Egg67}
	{Eggen}, O.~J., {Mathewson}, D.~S., \& {Serkowski}, K. 1967, \nat, 213, 1216
	
	\bibitem[{Evans {et~al.}(2012)Evans, Gehrz, Helton, Starrfield, Bode, Osborne,
		Banerjee, Ness, Walter, Woodward, Kuulkers, Eyres, Oliveira, Ashok, Krautter,
		O'Brien, Page, \& Rushton}]{Eva12}
	Evans, A., Gehrz, R.~D., Helton, L.~A., {et~al.} 2012, Monthly Notices of the
	Royal Astronomical Society: Letters, 424, L69
	
	\bibitem[{{Ferland} {et~al.}(2017){Ferland}, {Chatzikos}, {Guzm{\'a}n},
		{Lykins}, {van Hoof}, {Williams}, {Abel}, {Badnell}, {Keenan}, {Porter}, \&
		{Stancil}}]{Fer17}
	{Ferland}, G.~J., {Chatzikos}, M., {Guzm{\'a}n}, F., {et~al.} 2017, ArXiv
	e-prints [\eprint[arXiv]{1705.10877}]
	
	\bibitem[{{Godon} {et~al.}(2018){Godon}, {Sion}, {Williams}, \&
		{Starrfield}}]{God18}
	{Godon}, P., {Sion}, E., {Williams}, R., \& {Starrfield}, S. 2018, ArXiv
	e-prints [\eprint[arXiv]{1806.06059}]
	
	\bibitem[{{Goswami} \& {Karinkuzhi}(2013)}]{Gos13}
	{Goswami}, A. \& {Karinkuzhi}, D. 2013, \aap, 549, A68
	
	\bibitem[{{Goswami} {et~al.}(2010){Goswami}, {Kartha}, \& {Sen}}]{Gos10}
	{Goswami}, A., {Kartha}, S.~S., \& {Sen}, A.~K. 2010, \apjl, 722, L90
	
	\bibitem[{{Helton} {et~al.}(2010){Helton}, {Woodward}, {Walter},
		{Vanlandingham}, {Schwarz}, {Evans}, {Ness}, {Geballe}, {Gehrz},
		{Greenhouse}, {Krautter}, {Liller}, {Lynch}, {Rudy}, {Shore}, {Starrfield},
		\& {Truran}}]{Hel10}
	{Helton}, L.~A., {Woodward}, C.~E., {Walter}, F.~M., {et~al.} 2010, \aj, 140,
	1347
	
	\bibitem[{{Izzo} {et~al.}(2012){Izzo}, {Ederoclite}, {Della Valle}, {Mason},
		{Williams}, {Altamore}, {Cassatella}, {Gilmozzi}, {Patat}, {Schmidtobreick},
		{Selvelli}, {Tappert}, {Thater}, {Covone}, {Dall'Ora}, \& {Paolillo}}]{Izz12}
	{Izzo}, L., {Ederoclite}, A., {Della Valle}, M., {et~al.} 2012, \memsai, 83,
	830
	
	\bibitem[{{Joshi} {et~al.}(2014){Joshi}, {Banerjee}, \& {Ashok}}]{Jos14}
	{Joshi}, V., {Banerjee}, D.~P.~K., \& {Ashok}, N.~M. 2014, \mnras, 443, 559
	
	\bibitem[{{Livio} {et~al.}(1990){Livio}, {Shankar}, {Burkert}, \&
		{Truran}}]{Liv90}
	{Livio}, M., {Shankar}, A., {Burkert}, A., \& {Truran}, J.~W. 1990, \apj, 356,
	250
	
	\bibitem[{{Lloyd} {et~al.}(1997){Lloyd}, {O'Brien}, \& {Bode}}]{Llo97}
	{Lloyd}, H.~M., {O'Brien}, T.~J., \& {Bode}, M.~F. 1997, \mnras, 284, 137
	
	\bibitem[{{Mondal} {et~al.}(2018){Mondal}, {Anupama}, {Kamath}, {Das},
		{Selvakumar}, \& {Mondal}}]{Mon18}
	{Mondal}, A., {Anupama}, G.~C., {Kamath}, U.~S., {et~al.} 2018, \mnras, 474,
	4211
	
	\bibitem[{{Morisset}(2013)}]{Mor13}
	{Morisset}, C. 2013, {pyCloudy: Tools to manage astronomical Cloudy
		photoionization code}, Astrophysics Source Code Library
	
	\bibitem[{{Nelson} {et~al.}(2014){Nelson}, {Chomiuk}, {Roy}, {Sokoloski},
		{Mukai}, {Krauss}, {Mioduszewski}, {Rupen}, \& {Weston}}]{Nel14}
	{Nelson}, T., {Chomiuk}, L., {Roy}, N., {et~al.} 2014, \apj, 785, 78
	
	\bibitem[{{Patterson} {et~al.}(1998){Patterson}, {Kemp}, {Shambrook},
		{Thorstensen}, {Skillman}, {Gunn}, {Jensen}, {Vanmunster}, {Shugarov},
		{Mattei}, {Shahbaz}, \& {Novak}}]{Pat98}
	{Patterson}, J., {Kemp}, J., {Shambrook}, A., {et~al.} 1998, \pasp, 110, 380
	
	\bibitem[{{Patterson} {et~al.}(2017){Patterson}, {Oksanen}, {Kemp}, {Monard},
		{Rea}, {Hambsch}, {McCormick}, {Nelson}, {Allen}, {Krajci}, {Lowther},
		{Dvorak}, {Borgman}, {Richards}, {Myers}, {Harlingten}, \& {Bolt}}]{Pat17}
	{Patterson}, J., {Oksanen}, A., {Kemp}, J., {et~al.} 2017, \mnras, 466, 581
	
	\bibitem[{{Raj} {et~al.}(2018){Raj}, {Pavana}, {Kamath}, {Anupama}, \&
		{Walter}}]{Raj18}
	{Raj}, A., {Pavana}, M., {Kamath}, U.~S., {Anupama}, G.~C., \& {Walter}, F.~M.
	2018, \actaa, 68, 79
	
	\bibitem[{{Ramaprakash} {et~al.}(1998){Ramaprakash}, {Gupta}, {Sen}, \&
		{Tandon}}]{ram98}
	{Ramaprakash}, A.~N., {Gupta}, R., {Sen}, A.~K., \& {Tandon}, S.~N. 1998,
	\aaps, 128, 369
	
	\bibitem[{{Schaefer}(2018)}]{Sch18}
	{Schaefer}, B.~E. 2018, ArXiv e-prints [\eprint[arXiv]{1809.00180}]
	
	\bibitem[{{Schaefer} {et~al.}(2013){Schaefer}, {Landolt}, {Linnolt},
		{Stubbings}, {Pojmanski}, {Plummer}, {Kerr}, {Nelson}, {Carstens},
		{Streamer}, {Richards}, {Myers}, \& {Dillon}}]{Sch13}
	{Schaefer}, B.~E., {Landolt}, A.~U., {Linnolt}, M., {et~al.} 2013, \apj, 773,
	55
	
	\bibitem[{{Schaefer} {et~al.}(2010){Schaefer}, {Pagnotta}, \& {Shara}}]{Sch10}
	{Schaefer}, B.~E., {Pagnotta}, A., \& {Shara}, M.~M. 2010, \apj, 708, 381
	
	\bibitem[{{Schwarz}(2002)}]{Sch02}
	{Schwarz}, G.~J. 2002, \apj, 577, 940
	
	\bibitem[{{Schwarz} {et~al.}(2001){Schwarz}, {Shore}, {Starrfield},
		{Hauschildt}, {Della Valle}, \& {Baron}}]{Sch01}
	{Schwarz}, G.~J., {Shore}, S.~N., {Starrfield}, S., {et~al.} 2001, \mnras, 320,
	103
	
	\bibitem[{{Schwarz} {et~al.}(2007{\natexlab{a}}){Schwarz}, {Shore},
		{Starrfield}, \& {Vanlandingham}}]{Sch07a}
	{Schwarz}, G.~J., {Shore}, S.~N., {Starrfield}, S., \& {Vanlandingham}, K.~M.
	2007{\natexlab{a}}, \apj, 657, 453
	
	\bibitem[{{Schwarz} {et~al.}(2007{\natexlab{b}}){Schwarz}, {Woodward}, {Bode},
		{Evans}, {Eyres}, {Geballe}, {Gehrz}, {Greenhouse}, {Helton}, {Liller},
		{Lyke}, {Lynch}, {O'Brien}, {Rudy}, {Russell}, {Shore}, {Starrfield},
		{Temim}, {Truran}, {Venturini}, {Wagner}, {Williams}, \& {Zamanov}}]{Sch07b}
	{Schwarz}, G.~J., {Woodward}, C.~E., {Bode}, M.~F., {et~al.}
	2007{\natexlab{b}}, \aj, 134, 516
	
	\bibitem[{{Selvelli} {et~al.}(2008){Selvelli}, {Cassatella}, {Gilmozzi}, \&
		{Gonz{\'a}lez-Riestra}}]{Sel08}
	{Selvelli}, P., {Cassatella}, A., {Gilmozzi}, R., \& {Gonz{\'a}lez-Riestra}, R.
	2008, \aap, 492, 787
	
	\bibitem[{{Serkowski}(1974)}]{ser74}
	{Serkowski}, K. 1974, in Planets, Stars and Nebulae Studied with
	Photopolarimetry, Tucson, Univ. of Arizona Press, ed. T.~{Gehrels}, 135--174
	
	\bibitem[{{Shakhovskoy} {et~al.}(2017){Shakhovskoy}, {Antonyuk}, \&
		{Belan}}]{Sha17}
	{Shakhovskoy}, D.~N., {Antonyuk}, K.~A., \& {Belan}, S.~P. 2017, Astrophysics,
	60, 19
	
	\bibitem[{{Shara} {et~al.}(1989){Shara}, {Moffat}, {Williams}, \&
		{Cohen}}]{Sha89}
	{Shara}, M.~M., {Moffat}, A.~F.~J., {Williams}, R.~E., \& {Cohen}, J.~G. 1989,
	\apj, 337, 720
	
	\bibitem[{{Shara} {et~al.}(1997){Shara}, {Zurek}, {Williams}, {Prialnik},
		{Gilmozzi}, \& {Moffat}}]{Sha97}
	{Shara}, M.~M., {Zurek}, D.~R., {Williams}, R.~E., {et~al.} 1997, \aj, 114, 258
	
	\bibitem[{{Shore} {et~al.}(2013){Shore}, {Schwarz}, {De Gennaro Aquino},
		{Augusteijn}, {Walter}, {Starrfield}, \& {Sion}}]{Sho13}
	{Shore}, S.~N., {Schwarz}, G.~J., {De Gennaro Aquino}, I., {et~al.} 2013, \aap,
	549, A140
	
	\bibitem[{{Soker}(2015)}]{Soker15}
	{Soker}, N. 2015, \apj, 800, 114
	
	\bibitem[{{Sokoloski} {et~al.}(2013){Sokoloski}, {Crotts}, {Lawrence}, \&
		{Uthas}}]{Sok13}
	{Sokoloski}, J.~L., {Crotts}, A.~P.~S., {Lawrence}, S., \& {Uthas}, H. 2013,
	\apjl, 770, L33
	
	\bibitem[{{Surina} {et~al.}(2014){Surina}, {Hounsell}, {Bode}, {Darnley},
		{Harman}, \& {Walter}}]{Sur14}
	{Surina}, F., {Hounsell}, R.~A., {Bode}, M.~F., {et~al.} 2014, \aj, 147, 107
	
	\bibitem[{Svato{\v{s}}(1983)}]{svatovs1983intrinsic}
	Svato{\v{s}}, J. 1983, Astrophysics and Space Science, 93, 347
	
	\bibitem[{{Toraskar} {et~al.}(2013){Toraskar}, {Mac Low}, {Shara}, \&
		{Zurek}}]{Tor13}
	{Toraskar}, J., {Mac Low}, M.-M., {Shara}, M.~M., \& {Zurek}, D.~R. 2013, \apj,
	768, 48
	
	\bibitem[{{Vanlandingham} {et~al.}(2005){Vanlandingham}, {Schwarz}, {Shore},
		{Starrfield}, \& {Wagner}}]{Van05}
	{Vanlandingham}, K.~M., {Schwarz}, G.~J., {Shore}, S.~N., {Starrfield}, S., \&
	{Wagner}, R.~M. 2005, \apj, 624, 914
	
\end{thebibliography}
\end{document}